\documentclass{jfm}
\usepackage{graphicx}% Include figure files
\usepackage{dcolumn}% Align table columns on decimal point
\usepackage{bm}% bold math
\usepackage{booktabs}
\usepackage{xcolor}
\usepackage{dirtytalk}
\usepackage{centernot}
\usepackage{appendix}
\usepackage{amsmath, amssymb}
\usepackage{epstopdf}%So pdfs work for this template. 
\usepackage{float}%To force figures to output at a specific location

\newcommand{\boldu}{\textbf{u}}
\newcommand{\boldup}{\mathbf{u_p}}

\newcommand{\ez}{\hat{\textbf{e}}_z}

\newcommand{\rxmax}{r_{\max}}
\newcommand{\rmax}{r_{\rm{max}}}
\newcommand{\rsup}{r_{\rm{sup}}}
\newcommand{\usup}{u_{\rm{sup}}}

\newcommand{\rrms}{r_{\rm{rms}}}
\newcommand{\urms}{u_{\rm{rms}}}
\newcommand{\urmstot}{{U}_{\rm{rms}}}

\newcommand{\zmax}{z_{\rm{max}}}

\newcommand{\Tp}{{T_{\rm{p}}}}

\newcommand{\Pep}{{\rm{Pe_{p}}}}
\newcommand{\Pet}{{\rm{Pe}}_{T}}
\newcommand{\Reyn}{{\rm{Re}}}
\newcommand{\Rep}{{\rm{Re_{p}}}}

\def\div{{\mathbf \nabla \cdot}}

\def\grad{{\mathbf \nabla}}

\title{Preferential concentration in the particle-induced convective instability}
\author{Sara Nasab\aff{1}
  \corresp{\email{snasab@ucsc.edu}},
  Pascale Garaud\aff{1}, \corresp{\email{pgaraud@soe.ucsc.edu}}}
\affiliation{\aff{1} Department of Applied Mathematics, Baskin School of Engineering, University of California Santa Cruz, 1156 High Street, Santa Cruz, CA 95064, USA}
\begin{document}

\maketitle 

\begin{abstract}
Heavy particles in turbulent flows have been shown to accumulate in regions of high strain rate or low vorticity, a process otherwise known as preferential concentration. This can be observed in geophysical flows, and is inferred to occur in astrophysical environments, often resulting in rapid particle growth which is critical to physical processes such as rain or planet formation. Here we study the effects of preferential concentration in a two-way coupled system in the context of the particle-driven convective instability. To do so, we use Direct Numerical Simulations and adopt the two-fluid approximation. We focus on a particle size range for which the latter is valid, namely when the Stokes number is $\lesssim O(0.1)$. For Stokes number above $\sim 0.01$,  we find that the maximum particle concentration enhancement over the mean scales with the rms fluid velocity $u_{\rm{rms}}$, the particle stopping time $\tau_p$, and the particle diffusivity $\kappa_p$, as $u_{\rm{rms}}^2 \tau_p / \kappa_p$.  We show that this scaling can be understood from simple arguments of dominant balance. We also show that the typical particle concentration enhancement over the mean scales as $(u_{\rm{rms}}^2 \tau_p/ \kappa_p)^{1/2}$. We finally find that the probability distribution function of the particle concentration enhancement over the mean has an exponential tail whose slope scales as $(u_{\rm{rms}}^2 \tau_p / \kappa_p)^{-1/2}$. We apply our model to geophysical and astrophysical examples, and discuss its limitations. 
\end{abstract}

\section{Introduction}\label{sec:intro}
Preferential concentration is the tendency for heavy particles to accumulate in regions of high strain rate and low vorticity due to their inertia \citep{csanady1963turbulent, meek1973studies, maxey1987gravitational}. Investigations of the process date back to the 1980s and were performed using numerical experiments \citep{maxey1986gravitational, squires1991preferential, elghobashi1992direct} and laboratory experiments \citep{fessler1994preferential, kulick1994particle}. For comprehensive reviews of the topic, see for instance \citep{eaton1994preferential, crowe1996numerical, balachandar2010turbulent, monchaux2012analyzing} and references therein.

Today, thanks to progress in high-performance computing, Direct Numerical Simulations (DNSs) are a particularly convenient tool for quantifying preferential concentration in particle-laden flows. A variety of techniques can be used, which can be loosely classified into two distinct approaches: the Lagrangian-Eulerian and Eulerian-Eulerian approaches (see Sections \ref{subsec:lagrang}--\ref{subsec:twofluid} for more detail). The Lagrangian-Eulerian (LE) approach is named for the fact that the particles are evolved individually by integrating their equations of motion, while the carrier fluid is evolved on an Eulerian mesh. Various degrees of sophistication exist, depending on whether the particles are modeled realistically using, for instance, immersed boundary techniques \citep{mittal2005immersed}, or in a simplified way, as point particles \citep{toschi2009LagrangianReview}. In the latter case, particles can either be passively advected, or can react back on the fluid through drag.  When particles are modeled exactly, the LE approach is capable of modeling particle-particle interactions, such as collisions. Otherwise, these interactions must be accounted for using simplified parameterizations instead. However, as the number of particles increases, the computational cost can be expensive. In the Eulerian-Eulerian (EE) approach by contrast, the particles are treated as a continuum field with its own momentum and mass conservation laws, which are evolved on an Eulerian mesh \citep{elghobashi1994predicting, crowe1996numerical, morel2015mathematical}. Within the EE framework, various levels of approximation exist depending on the size of the particles, ranging from the so-called equilibrium Eulerian limit \citep{ferry2002, ferry2003improveEE} in which particle inertia is neglected, to the two fluid limit \citep{elghobashi1983two, druzhinin1998direct} which is generally more valid for somewhat larger particles. 

In regions of the fluid that experience strong local enhancement in the particle number density, increased collision rates can result in rapid particle growth \citep{cuzzi2001size}. As such, preferential concentration is thought to play an important role in controlling the size distribution function of particles suspended in turbulent fluids. Prior works have focused on certain aspects of preferential concentration such as the enhancement of the particle settling velocity \citep{aliseda2002effect, maxey1987gravitational, wang1993settling, mei1994effect, yang1998role, bosse2006small}, the resulting geometry of the dense particle clusters \citep{cuzzi2001size, monchaux2010preferential, goto2006self}, and the underlying mechanisms responsible for inertial clustering of particles \citep{raju1995accumulation, obligado2014preferential, goto2008sweep, coleman2009unified}. Preferential concentration likely plays a key role in the warm rain formation in clouds  \citep{pinsky2002effects, falkovich2002acceleration, riemer2005droplets}, protoplanetary disks \citep{klahr1997particle, chambers2010planetesimal, cuzzi2008toward}, estuaries \citep{eisma1991particle, VOULGARIS20041659}, and industrial applications such as sprays \citep{cao2000piv, vie2015analysis}. In all of these examples, some of the key questions that remain to be answered are: (1) What is the maximum particle concentration enhancement that can be achieved anywhere in the fluid? (2) What is the typical probability distribution of the volume density of particles? And, (3) how do these quantities depend on the turbulent properties of the carrier flow? 

While these questions have been primarily investigated in forced turbulent flows so far \citep{squires1991preferential, eaton1994preferential, bosse2006small}, they have not been studied extensively to our knowledge in the context of particle-induced buoyancy instabilities (e.g. convective or Rayleigh-Taylor). Such instabilities are particularly relevant in particle-laden turbidity currents, which play an important role in sediment transport \citep{meiburg2010turbidity}. Although most research to date on particle-laden buoyancy-driven flows has been performed using in-situ or laboratory experiments \citep{hoyal1999settling, maxworthy1999dynamics, parsons2001hyperpycnal, voltz2001rayleigh}, numerical experiments have only recently begun to be used in this context. The focus of these numerical studies can be categorized into two groups: (1) numerical tests, in which various formalisms (i.e. LE versus EE) are compared to one another \citep{chou2014towardpart1, chou2014towardpart1, chou2016numerical}, and (2) application-driven studies, that investigate for instance how the rate of sedimentation is influenced by particle properties. It was shown that both particle size and particle volume fraction can control the resulting modes of instability (i.e. leaking, fingering, stable settling modes) from the initial RT instability configuration, affecting the subsequent evolution of the sedimentation process \citep{burns2012sediment, burns2015sediment, shao2017numerical}. However, numerical investigations whose primary focus is on preferential concentration in the particle-driven convective instability, specifically for two-way coupled systems have not been performed. 

In this paper, we therefore study preferential concentration in the two-way coupled two-fluid formalism using DNSs of particle-driven convective instabilities. Section \ref{sec:twofluid} describes the two-fluid formalism. In Section \ref{sec:model}, we introduce our model setup and its governing equations. In Section \ref{sec:num_sims}, we present the results of the DNSs and investigate how certain parameters influence preferential concentration and the underlying turbulence. In Section \ref{sec:analysis}, we present a predictive model that captures maximum particle concentration enhancement as a function of time and space. In Section \ref{sec:pdf}, we look at the probability distribution function (PDF) of the relative particle concentration. Section \ref{sec:summary} summarizes our results and discusses them in the context of geophysical and astrophysical applications of particle-laden flows.

\section{Two-fluid formalism}\label{sec:twofluid}
The two-fluid formalism for particle-laden flows can be derived starting from the Lagrangian-Eulerian formalism by locally averaging the particle properties to obtain the continuum density and momentum conservation equations. This essentially follows the derivation of Ishii and Mishima  \citep{ishii1984two} (see also \citep{ishii2010thermo, delhaye1976averaging}). The formalism has been widely used within the astrophysics community for studying protoplanetary disks \citep{youdin2005streaming, nakagawa1986settling}, as well as in studies related to sediment transport in rivers and oceans \citep{hsu2004two, bakhtyar2009two, revil2013two}, for instance. 

For simplicity in this work, we focus on particulate flows in which the particle solid density $\rho_s$ is much larger than the mean density of the fluid $\rho_f$, such as droplets or aerosols in the atmosphere or dust in accretion disks. We also assume that the particles are spherical, monodisperse, and dilute (ensuring that particle-particle collisions do not dominate the particle evolution equations). 

\subsection{Lagrangian formalism}\label{subsec:lagrang}
Under the above assumptions, we can model the motion of a single particle interacting with the fluid through Stokes drag by solving the coupled ordinary differential equations  
\begin{equation}\label{eqn:singleparticle}
	\frac{d \mathbf{x_p}}{dt} = 	\boldup \; \; \; \; \; \mbox{and} \; \; \; \; \; \frac{d \boldup}{dt} = \frac{\boldu(\mathbf{x_p}) - \boldup}{\tau_p} + \mathbf{g},
\end{equation}
where $\mathbf{x_p}$ is the position of the particle, $\boldup$ is its velocity, $\boldu(\mathbf{x_p})$ is the fluid velocity at $\mathbf{x_p}$, $\mathbf{g} = -g\ez$ is gravity, and $\tau_p$ is the particle stopping time. In \eqref{eqn:singleparticle}, we have assumed that the reduced mass (which would normally multiply $\mathbf{g}$)  is approximately 1 since $\rho_s \gg \rho_f$. We have also neglected other effects such as the Basset history and Saffman lift terms for the same reason \citep{maxey1983equation}.

To model a collection of $N_p$ monodisperse particles using the LE approach, \eqref{eqn:singleparticle} is integrated separately for each particle in the fluid:
\begin{align}
	\label{Lagrangian}
	\frac{d \mathbf{x}_{p,i}}{dt} = \boldu_{p,i}  \; \; \; \; \; \mbox{and} \; \; \; \; \; \frac{d \textbf{u}_{p,i}}{dt} = \frac{\boldu(\mathbf{x}_{p,i}) - \textbf{u}_{p,i}}{\tau_{p}} + \textbf{g} \hspace{.2in} \text{for } i = 1, ..., N_p,
\end{align}
where $\mathbf{x}_{p,i}$ and $\textbf{u}_{p,i}$ are the position and velocity of the $i$th particle, respectively. The back reaction of the particles on the fluid is accounted for by adding a mean local drag force $\mathbf{F}_p$ in the Navier-Stokes equation (shown here in the limit of the Boussinesq approximation \citep{boussinesq1903comptes, spiegel1960boussinesq}): 
\begin{equation}
	\label{eqn:NS} \rho_f \bigg( \frac{\partial \boldu}{\partial t} + \boldu \cdot \grad \boldu \bigg) = -\grad p + \rho \mathbf{g} + \rho_f \nu \grad^2 \boldu + \mathbf{F}_p,
\end{equation}
where $\rho$ is the density deviation away from the mean fluid density $\rho_f$, $p$ is the pressure, $\nu$ is the kinematic velocity of the fluid, and $\mathbf{F}_p$ is explicitly defined as 
\begin{equation}
	\label{eqn:dragLag} \mathbf{F}_p(\mathbf{x}) = -\sum^{N_p}_{i=1}  \frac{m_p}{v_\epsilon} \frac{\boldu(\mathbf{x}_{p,i}) -\textbf{u}_{p,i}}{\tau_{p}}  H(\epsilon-|\mathbf{x}_{p,i} - \mathbf{x}|),
\end{equation}
where $H$ is the Heaviside function, $m_p$ is the particle mass, and $v_\epsilon$ is the volume of a sphere of radius $\epsilon$. The averaging radius $\epsilon$ is typically chosen to be one grid cell in numerical computations. Equations \eqref{Lagrangian}- \eqref{eqn:dragLag}, together with the fluid incompressibility condition $\div \boldu = 0$, form the Lagrangian-Eulerian equations. These can now be statistically averaged using methods motivated from kinetic theory to derive the two-fluid formalism. 

\subsection{Two-fluid formalism}\label{subsec:twofluid}
We first define the local mass density of particles $\rho_p$ and corresponding velocity $\boldup$, averaged in a small volume centered around the position $\mathbf{x}$ as
\begin{equation}
	\label{eqn:continuum}\rho_p = \frac{m_p}{v_\epsilon } \sum^{N_p}_{i=1}  H(\epsilon - |\mathbf{x}_{p,i} - \mathbf{x}|) , \hspace{.3in}  \boldup(\mathbf{x}) = \frac{1}{v_\epsilon} \sum^{N_p}_{i=1} \textbf{u}_{p,i} H(\epsilon - |\mathbf{x}_{p,i} - \mathbf{x}|).
\end{equation}
Applying this average to the particle evolution equations in \eqref{Lagrangian} (as done in \citep{ishii1984two}, for instance), we approximately get 
\begin{equation}
	\frac{D_p \boldup}{D_p t} = \frac{\boldu- \boldup}{\tau_p} + \mathbf{g}  + ... 
\end{equation}
where $D_p/D_p t = \partial/\partial t + \boldup \cdot \nabla$  is the derivative following the mean particle velocity.  The evolution equation for the particle density can be obtained by mass conservation to be  
\begin{equation}  
	\frac{\partial \rho_p}{\partial t} + \div (\rho_p \boldup) = ... 
\end{equation}

In both equations, dots on the right hand side result from three possible sources: (1) dispersion in both mass and momentum conservation equations due to the fact that $\boldu_{p,i} \neq \boldup$; (2) unaccounted for interactions of the particles with the fluid, which include Brownian motions for very small particles, and self-interaction of the particle with its own wake if the latter is not perfectly modeled by the Stokes solution; and (3) long-range interactions of particles with one another due to each other's wakes. Aside from Brownian motions, these terms are generally very difficult to model, leading to strong anisotropic dispersion, and likely to depend nonlinearly on the mean particle density and velocity. 

In what follows, we will model these terms for simplicity as $\nu_p \nabla^2 \boldup$ in the momentum equation and $\kappa_p \nabla^2 \rho_p$ in the density equation, so
\begin{align}
	&\label{eqn:2fluid}\frac{\partial \boldup}{\partial t} + \boldup \cdot \grad \boldup + \frac{\boldup - \boldu}{\tau_p} - \textbf{g} = \nu_p \grad^2 \boldup, \\
	& \label{eqn:2fluid2} \frac{\partial \rho_p}{\partial t} + \div (\rho_p \boldup) = \kappa_p \grad^2 \rho_p.
\end{align}
These terms are included to stabilize the numerical scheme in the DNSs, although they are also physically motivated in the limit where Brownian motion is the dominant source of dispersion. Note that we anticipate the two-fluid approach to break down when the Stokes number (the ratio of the stopping time to the eddy turnover time) approaches unity, in which case the particles become uncorrelated with the fluid and therefore also with one another \citep{shotorban2006particle}. When this happens, the mean particle velocity $\boldup$ is no longer a good approximation for each individual particle velocity, and the averaging procedure becomes meaningless. 

To couple the particle and fluid evolution equations, note that the drag term in the continuum limit in (\ref{eqn:NS}) becomes
\begin{equation}
	\mathbf{F}_p(\mathbf{x}) = \rho_p(\mathbf{x}) \frac{\boldup(\mathbf{x}) - \boldu(\mathbf{x})}{\tau_p}, 
\end{equation}
so the two-way coupled equations are
\begin{align}\label{eqn:twofluid1}
&\rho_f \bigg( \frac{\partial \boldu}{\partial t} + \boldu \cdot \grad \boldu \bigg) = -\grad p + \rho \textbf{g}+\rho_p \frac{\boldup - \boldu}{\tau_p}  + \rho_f \nu \grad^2 \boldu ,\\
\label{eqn:twofluid2}
	&\frac{\partial \boldup}{\partial t} + \boldup \cdot \grad \boldup + \frac{\boldup - \boldu}{\tau_p} - \textbf{g} = \nu_p \grad^2 \boldup, \\
	\label{eqn:twofluid3}
	& \frac{\partial \rho_p}{\partial t} + \div (\rho_p \boldup) = \kappa_p \grad^2 \rho_p, \\ 
	\label{eqn:twofluid4}
	&\div \boldu = 0. 
\end{align}

For smaller particles that are well-coupled to the fluid, the two-fluid formalism recovers the equilibrium Eulerian formalism in which particle inertia is negligible. We can demonstrate this by taking the formal limit $\tau_p \rightarrow 0$ to obtain: 
\begin{align}
   	& \boldup = \boldu - w_s \ez, \label{eqn:ee1} \\
	& \rho_f \bigg( \frac{\partial \boldu}{\partial t} + \boldu \cdot \nabla \boldu \bigg) = -\nabla p + (\rho + \rho_p) \mathbf{g} + \rho_f \nu \nabla^2 \boldu, \label{eqn:ee2} \\
	& \frac{\partial \rho_p}{\partial t} + (\boldu - w_s \ez) \cdot \nabla \rho_p = \kappa_p \nabla^2 \rho_p, \label{eqn:ee3} \\
	& \div \boldu = 0 \label{eqn:ee4},
\end{align}
where the settling velocity $w_s$ is related to the stopping time and gravity via $w_s = \tau_p g$. The particle velocity $\boldup$ is now determined by the carrier fluid velocity and the particle settling velocity. Compared to the two-fluid formalism, we see that $\div \boldup \equiv 0$; thus, the particle concentration is solely advected by the carrier flow. As a result, preferential concentration cannot be captured by the equilibrium Eulerian approach \citep{maxey1983equation}.
%---

\section{The Model}\label{sec:model}
\subsection{Model set-up}
We investigate particle-driven convective instabilities in a dilute suspension using the two-fluid equations. For convenience, we rescale the particle density with the mean density of the fluid, which defines $r = \rho_p/\rho_f$. Having assumed that $\rho_s \gg \rho_f$, it is still possible to have $r$ of order unity even though the volume fraction of particles is assumed to be very small. We assume that the carrier fluid has a constant stable background temperature gradient $T_{0z} > 0$ in the vertical direction, with the background temperature profile given by $T_0(z) = T_m + zT_{0z}$. This assumption was originally motivated by applications in which the carrier fluid is typically stratified, such as in warm clouds or rivers, but does not directly impact the results presented in this paper. Perturbations in the density of the carrier fluid $\rho$ are caused by temperature fluctuations $T$ around that background profile, and are related via $\rho/\rho_f = -\alpha T$, where $\alpha = -\rho_f^{-1} (\partial \rho / \partial T)$. \\

In the limit of the Boussinesq approximation, the governing dimensional equations are then
\begin{align}
	&\frac{\partial \boldu}{\partial t} + \boldu \cdot \grad \boldu = -\frac{\grad p}{\rho_f} + \alpha g T \ez  + r \frac{\boldup -\boldu}{\tau_p} + \nu \grad^2 \boldu, \label{eqn:dim1} \\
	& \frac{\partial \boldup}{\partial t} + \boldup\cdot\grad\boldup = \frac{\boldu - \boldup}{\tau_p} + \mathbf{g} + \nu_p \grad^2 \boldup, \label{eqn:dim2}\\
	&\frac{\partial r}{\partial t} + \grad \cdot (\boldup r) =\kappa_p \grad^2 r, \label{eqn:dim3}\\
	&\frac{\partial T}{\partial t} + \boldu \cdot \grad T + w T_{0z} = \kappa_T \grad^2T, \label{eqn:dim4}\\
	& \div \boldu = 0 \label{eqn:dim5},
\end{align}
where  $\boldu = (u,v,w)$  and $\boldup = (u_p, v_p, w_p)$. 

Using this system of equations, we shall study the evolution of the relative particle density $r$. To do so in the context of the convective instability, we start with initial conditions that take the form of a Gaussian profile of amplitude $r_0$ and width $\sigma$:
\begin{equation}
	r(x,y,z,0) = r_0 \exp\bigg[ \frac{(z-L_z/2)^2}{2\sigma^2} \bigg],
\end{equation} 
 to which low amplitude random fluctuations are added, and where $L_z$ is the height of the computational domain. The initial particle velocity is set to be the particle settling velocity $w_s$, while the carrier fluid is initialized with zero velocity. 

\subsection{Non-dimensionalization}\label{subsec:nondim}
We define the units of length $[l]$, relative particle concentration $[r]$, and temperature $[T]$ as
\begin{align}\label{nondimensionalization}
	[l] = \sigma, \hspace{.2in} [r] = r_0, \hspace{.2in} [T] = \sigma T_{0z}.
\end{align}
We can define a characteristic velocity for the fluid by identifying its kinetic energy with an estimate of the potential energy of the unstable particle density distribution:  
\begin{equation}\label{eqn:u}
	[u] = \sqrt{r_0 g \sigma}.
\end{equation}
The characteristic distance and velocity can finally be used to construct a typical convective eddy turnover time
\begin{equation}
	[t] = \bigg( \frac{\sigma}{r_0 g} \bigg) ^ {1/2}.
\end{equation}
Thus, the non-dimensional equations are: 
\begin{align}
	&\frac{\partial \boldu}{\partial t} + \boldu \cdot \grad \boldu = -\grad p + R_{\rho} T {\ez} + \frac{1}{\Reyn} \grad^2 \boldu + {r_0} \bigg(r \frac{\boldup -\boldu}{T_{\rm{p}}} \bigg), \label{eqn:nondim1} \\
& \frac{\partial \boldup}{\partial t} + \boldup\cdot\grad\boldup= \frac{\boldu - \boldup}{T_{\rm{p}}} - \frac{1}{r_0} {\ez} + \frac{1}{\Rep} \grad^2 \boldup, \label{eqn:nondim2} \\
  &\frac{\partial r}{\partial t} + \grad \cdot (\boldup r) = \frac{1}{\Pep} \grad^2 r, \label{eqn:nondim3} \\
  &\frac{\partial T}{\partial t} + \boldu \cdot \grad T + {w}  = \frac{1}{\Pet} \grad^2 T, \label{eqn:nondim4} \\
	& \grad \cdot \boldu = 0, \label{eqn:nondim5}
\end{align}
where all the variables $(\boldu, \boldup, p, r, T)$ are from here on implicitly non-dimensional, and where the dimensionless parameters are defined as: 
\begin{align*}
	&R_{\rho} = \frac{\alpha \sigma T_{0z}}{r_0}  &\Reyn = \frac{(r_0 g)^{1/2} \sigma^{3/2}}{\nu}  &\hspace{.2in} &\Pet =  \frac{(r_0 g)^{1/2} \sigma^{3/2}}{\kappa_T}\\
    &T_{\rm{p}} = \tau_p \bigg(\frac{r_0 g}{\sigma}\bigg) ^ {1/2}  &{\Rep} =  \frac{(r_0 g)^{1/2} \sigma^{3/2}}{\nu_p}  &\hspace{.2in} &\Pep =  \frac{(r_0 g)^{1/2} \sigma^{3/2}}{\kappa_p} \\
&W_s = \frac{\Tp}{r_0}.
\end{align*}

Four of these parameters describe diffusive effects: a Reynolds number for the fluid $\Reyn$, a Reynolds number for the particles $\rm Re_p$, the particle P$\rm{\acute{e}}$clet number $\rm Pe_p$, and the temperature P$\rm{\acute{e}}$clet number ${\rm Pe}_T$. In the fluid momentum equation, $R_{\rho}$ is the density ratio, defined by analogy with double-diffusive systems to be the ratio of the density gradient due to temperature stratification $\alpha T_{0z}$ to the density gradient due to particle stratification, here estimated as $r_0 / \sigma$. In addition, $T_{\rm{p}}$ is the non-dimensional stopping time, and $W_s$ is the non-dimensional settling velocity of the particles. Note that our non-dimensionalization defines $T_{\rm{p}}$ as the ratio of the particle stopping time to the estimated turnover time of the layer-scale eddies. Thus, by construction, $T_{\rm{p}}$ is an estimate of the Stokes number $\rm{St}$ of the convectively turbulent flow.

We define the non-dimensional total density (i.e. consisting of the fluid and the particles) in the system as 
\begin{equation}
	{\rho}_{tot} = \bigg( \frac{1}{\alpha \sigma T_{0z}} - z - T \bigg) + \frac{r}{R_\rho}, 
\end{equation}
so the non-dimensional total background density gradient is 
\begin{equation}
	\frac{d {\rho}_{tot}}{d{z}} =  - \bigg( 1 + \frac{dT}{dz} \bigg) + \frac{1}{R_\rho} \frac{d r}{d z}.
\end{equation}
The total density gradient controls the development of the convective instability and, as shown above, is the sum of the density gradient due to the temperature stratification and the density gradient due to the particle stratification. At time $t = 0$, the non-dimensional initial condition for the particle concentration is 
\begin{equation}\label{eqn:r_init}
		r(x,z,0) = e^{-(z-L_z/2)^2/2}.
\end{equation}
The particle density gradient is most unstable at the lower inflection point of the Gaussian ($z = z_i$), where it is equal to 
\begin{equation}
  \frac{dr}{dz}\Bigg| _{z=z_i} = e^{-1/2}.
\end{equation}
Thus, the total density gradient at the lower inflection point $z = z_i$ at $t = 0$ is
\begin{equation}
	\frac{d {\rho}_{tot}}{d{z}}\Bigg|_{z = z_i, t=0}  =  - 1 + \frac{e^{-1/2}}{R_\rho}.
\end{equation}
Using this information, we define a Rayleigh number as
\begin{equation}\label{eqn:Ra_dimensional}
  	{\rm{Ra}} = \bigg(\frac{1}{\rho_f} \frac{d\rho_{tot}}{dz} \Big|_{z=z_i}\bigg) \frac{g\sigma^4}{\kappa_p \nu},
  	\end{equation}
where all the quantities on the right-hand side are dimensional. We can then express \eqref{eqn:Ra_dimensional} in terms of the previously defined dimensionless parameters as
\begin{equation}\label{eqn:Ra_nondimensional}
	{\rm{Ra}} = \bigg(\frac{e^{-1/2}}{R_\rho} - 1\bigg) \Reyn \Pep.
\end{equation}

To ensure that overturning convection (rather than double-diffusive instabilities) takes place in all that follows, we set $R_{\rho} = 0.5 < e^{-1/2}$.  We shall then vary $\rm{Ra}$ by varying either $\Reyn$ or $\Pep$, ensuring in all cases that $\rm{Ra}$ is sufficiently large for turbulent convection to take place. Finally, the Prandtl number will be fixed and equal to one for the flow to be fairly turbulent for all simulations. This choice fixes the relationship between $\Reyn$ and $\Pet$: 
\begin{equation}
	{\rm{Pr}} = \frac{\Pet}{\Reyn} \equiv 1.
\end{equation}

\section{Numerical Simulations}\label{sec:num_sims}
Since our goal is to characterize preferential concentration of the particles by the fluid, which is an inherently nonlinear phenomenon, we must use DNSs. In order to do so, we use the triply periodic pseudospectral PADDI code \citep{stellmach2011dynamics, traxler2011dynamics} which has been extensively used to study fingering as well as a number of astrophysical instabilities such as semi-convection and shear \citep{moll2016new, garaud2016turbulent}. A slightly modified version of the code was also used to study fingering convection in the equilibrium Eulerian regime \citep{reali2017layer}. We have modified the PADDI code further by adding a particle field which evolves according to the two fluid equations \eqref{eqn:dim1} - \eqref{eqn:dim5}, and refer to the new version of this code as PADDI-2F. Salient properties of PADDI, as well as the modifications made to include the two-fluid formalism, are briefly described in Appendix \ref{app:num_sims}.

In what follows, we present 2D and 3D simulations with specifications listed in Table \ref{table:sims}. The size of the computational domain is selected based on the following considerations: (1) since the code is triply-periodic, the domain height must be sufficiently large to avoid unphysical interactions between the particles that leave the domain at the bottom and re-enter it at the top. With that in mind, we present simulations with height ranging from $L_z = 10$ to $L_z = 20$. (2) The domain width must be chosen to be large enough to ensure that there are enough convective eddies in the horizontal direction to have meaningful statistics. In all the simulations presented below $L_x = 10$, and for 3D simulations, we further choose $L_y = 2$.

\subsection{Two-fluid code validation against Eulerian simulations}\label{subsec:code_validation}
To validate the PADDI-2F code, we begin by comparing a two-fluid simulation with low $T_{\rm{p}}$ solving equations \eqref{eqn:nondim1} -- \eqref{eqn:nondim5} with that of an equilibrium Eulerian simulation solving \eqref{eqn:ee1} -- \eqref{eqn:ee4} (used in \citep{reali2017layer}). In both codes, we set $W_s = 0.1, R_\rho = 0.5, \Reyn = 1000, {\rm{Pr}} = 1$, ${\Pet} = \rm{Pr} {\Reyn}$, and $\Pep = 1000$ (corresponding to $\rm{Ra} \approx 10^6$); for the two-fluid simulation, we additionally set the particle stopping time to be $T_{\rm{p}} = 0.005$, which should be sufficiently small to be in the limit where the equilibrium Eulerian formalism is valid. We first compare the two codes using 2D simulations (see Section \ref{subsec:comp_2D_3D} for a comparison of 2D vs. 3D simulations). We set the resolution of the 2D runs to be $768 \times 1536$ equivalent grid points in the $x-$ and $z-$ directions, respectively, and set the domain width and height as $L_x = 10$ and $L_z = 20$. 

\begin{figure} 
  \centerline{\includegraphics[width=.6\pdfpagewidth]{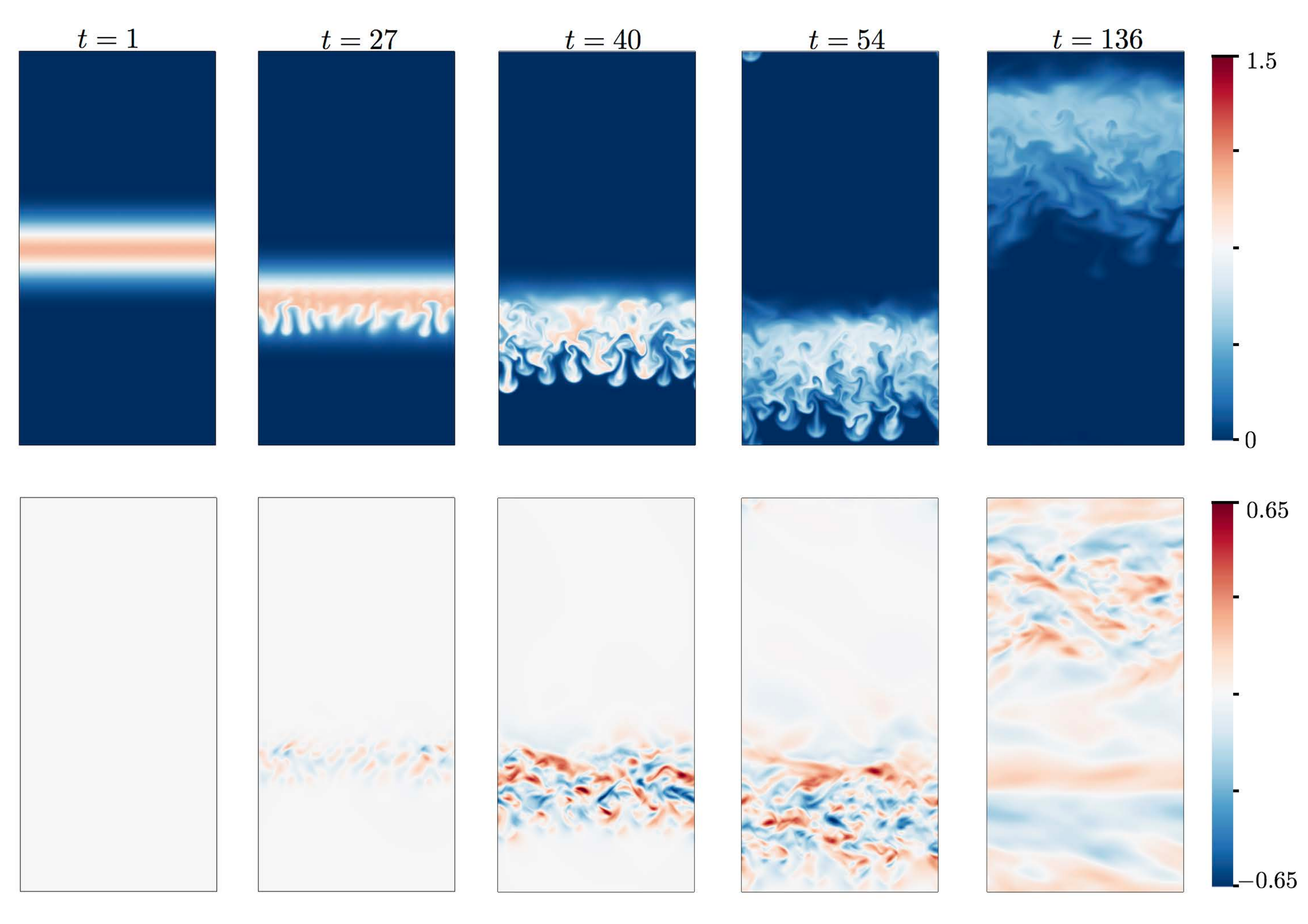}}% Images in 100% size
  \caption{Snapshots of the particle concentration $r$ (top row) and the horizontal component of the fluid velocity $u$ (bottom row) at various times in a two-fluid simulation with $T_{\rm{p}} = 0.005$, $W_s = 0.1, R_\rho = 0.5, \Reyn = 1000$, ${\rm{Pr}} = 1$.}
\label{sim2_snapshots}
\end{figure}

In the snapshots presented in Figure \ref{sim2_snapshots}, we see the evolution of the particle concentration and the horizontal component of the fluid velocity $u$ in the two-fluid simulation. Snapshots of the Eulerian simulation (not shown) taken at the same times look very similar to the two-fluid simulation (bearing in mind the chaotic nature of the system). The initially unstable total density stratification $\rho_{tot} (z, 0)$ drives the growth of convective eddies, which become visible in the second snapshot ($t = 27$). The particle layer then rapidly spreads vertically under the effect of turbulent mixing in the third snapshot ($t = 40$), reducing the unstable particle gradient. Although there are horizontal inhomogeneities in the particle concentration, these remain small compared with the horizontal mean. In particular, $r$ never exceeds the initial maximum value of one, consistent with the expected properties of an advection-diffusion equation when $\div \boldup \simeq 0$.  This shows qualitatively that for sufficiently small $T_{\rm{p}}$, the two-fluid simulation recovers behavior expected in the absence of particle inertia.

We now compare these simulations more quantitatively by examining the behavior of both the particle concentration and the fluid velocity. In order to do so, we define a number of diagnostic quantities (for convenience listed in Table \ref{table:defns}). We first define the maximum particle concentration and maximum horizontal fluid velocity in the domain at any point in time as 
\begin{equation}\label{rsup}
	r_{\rm{sup}}(t) = \max_{x,z} r(x,z,t) \text{ \hspace{.1in} and \hspace{.1in} } \usup(t) = \max_{x,z} {u(x,z,t)}.
\end{equation}
 We have selected to look at the behavior of the horizontal component of the velocity, rather than its vertical component or total amplitude, because it is not directly influenced by the particle settling motion.

In order to study the evolution of the bulk of the particle layer, we next define the horizontally averaged particle concentration profile $\overline{r}(z,t)$, where the overbar denotes a horizontal average, as in $\overline{q}(z,t) = \frac{1}{L_x} \int q(x,z,t) dx$ for any quantity $q$. The quantity $\bar{r}$ can be compared to the corresponding analytical expression obtained when the particles evolve purely diffusively, namely when
\begin{equation}\label{eqn:diff}
	\frac{\partial r_{\rm{diff}}}{\partial t} - W_s \frac{\partial r_{\rm{diff}}}{\partial z} = \frac{1}{\Pep} \grad ^2 r_{\rm{diff}}.
\end{equation}
The solution of \eqref{eqn:diff} in an infinite domain with initial condition given by \eqref{eqn:r_init} is
\begin{equation}\label{eqn:diff_sol}
	r_{\rm{diff}} (z,t) = \frac{1}{\sqrt{\frac{2}{\Pep} t + 1}} \exp\bigg[ -\frac{ [ z-(L_z/2-W_s t)]^2}{2[(2/\Pep)t + 1]} \bigg].
\end{equation}
As long as $2 t/\Pep \ll L_z$, this solution is also a good approximation to the diffusive solution in the periodic domain.

We also extract the maximum value of $\bar{r}$ at time $t$, which occurs at the height $z = z_{\max}(t)$
\begin{equation}\label{eqn:rbarstar}
	\bar{r}^*(t) = \bar{r}(z_{\max},t) = \max_{z} \bar{r} (z,t). 
\end{equation}
In what follows, the asterisk will always indicate a quantity measured at the position $z_{\max}(t)$. We can compare $\rsup$ and $\bar{r}^*$ to the maximum value of the diffusive solution, namely 
\begin{equation}\label{eqn:rmaxdiff}
	r_{\rm{diff, sup}}(t) = \max_z r_{\rm{diff}} (z,t) = \frac{1}{\sqrt{\frac{2}{\Pep} t +1}}.
\end{equation}

Finally, we define the root mean square of the $x-$component of the fluid velocity at a particular height $z$ and time $t$, expressed as 
\begin{equation}
	\urms(z,t) = \big[ \hspace{1 pt} \overline{u(x,z,t)^2} \hspace{1 pt} \big]^{1/2}.
\end{equation}
We can study turbulence in the \textit{bulk} of the particle layer over time by extracting the corresponding value of $\urms$ at the position $z = z_{\max}$, defined by
\begin{equation}\label{eqn:urmsstar}
	\urms^*(t) = \urms(z_{\max}, t). 
\end{equation}

\begin{figure} 
\caption{Low $T_{\rm{p}} = 0.005$ two-fluid simulation versus an equilibrium Eulerian simulation with $W_s = 0.1, R_\rho = 0.5, \Reyn = 1000, \Pep = 1000$, and ${\rm{Pr}} = 1$, comparing various diagnostics of the particle concentration (a) and of the horizontal component of fluid velocity (b).}
\centering 
\label{sim2_vs_Eulerian}
\vspace{.3cm}
{\includegraphics[width=0.75\pdfpagewidth]{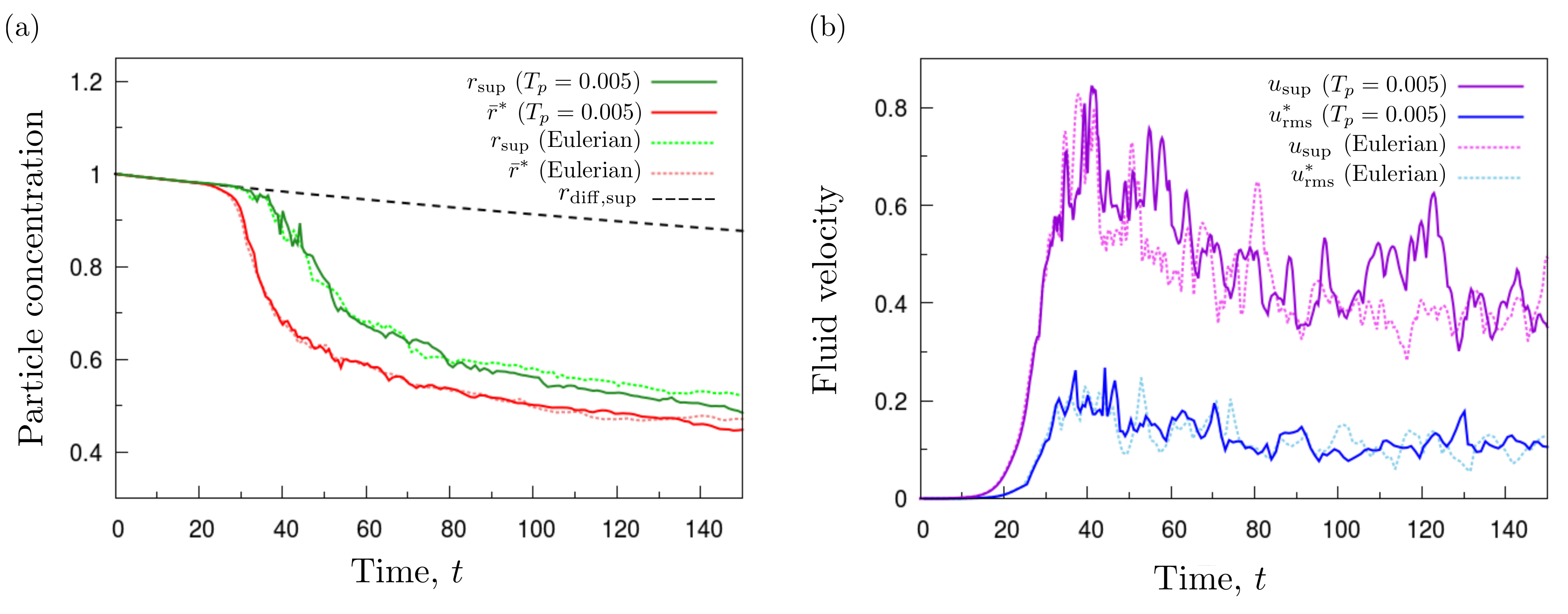}}
\end{figure}

Figure \ref{sim2_vs_Eulerian} shows a comparison of $\bar{r}^*$ and $\rsup$ for the particle concentration (Figure \ref{sim2_vs_Eulerian}a) and $\urms^*$ and $\usup$ for the fluid velocity (Figure \ref{sim2_vs_Eulerian}b) for both the two-fluid and equilibrium Eulerian simulations. Notably, we see that all the measured quantities are statistically consistent with one another in the two cases, verifying that the two-fluid formalism recovers the equilibrium Eulerian formalism for small $T_{\rm{p}}$. At early times ($t=0 - 25$) prior to the development of the convective instability, $\bar{r}^*$ and $\rsup$ follow the purely diffusive solution $r_{\rm{diff, sup}}$, shown as the black dotted line given by \eqref{eqn:rmaxdiff}.  Later, we see that $\bar{r}^*$ and $\rsup$ decrease rapidly (at times $t = 30 - 50$), then more slowly again after $t = 60$. During that time $\bar{r}^*$ and $\rsup$ roughly decay at the same rate. 

\begin{figure}
\caption{Evolution of the mean particle concentration $\bar{r}$ and of the rms fluid velocity  $\urms$ (defined in the main text) profiles for the two-fluid $T_{\rm{p}} = 0.005$ (red) and equilibrium Eulerian simulations (green). The black dotted Gaussian curve (first row) represents the purely diffusive solution \eqref{eqn:diff_sol} for comparison.}
\centering 
\label{sims_evolution}
\vspace{.3cm}
{\includegraphics[width=0.6\pdfpagewidth]{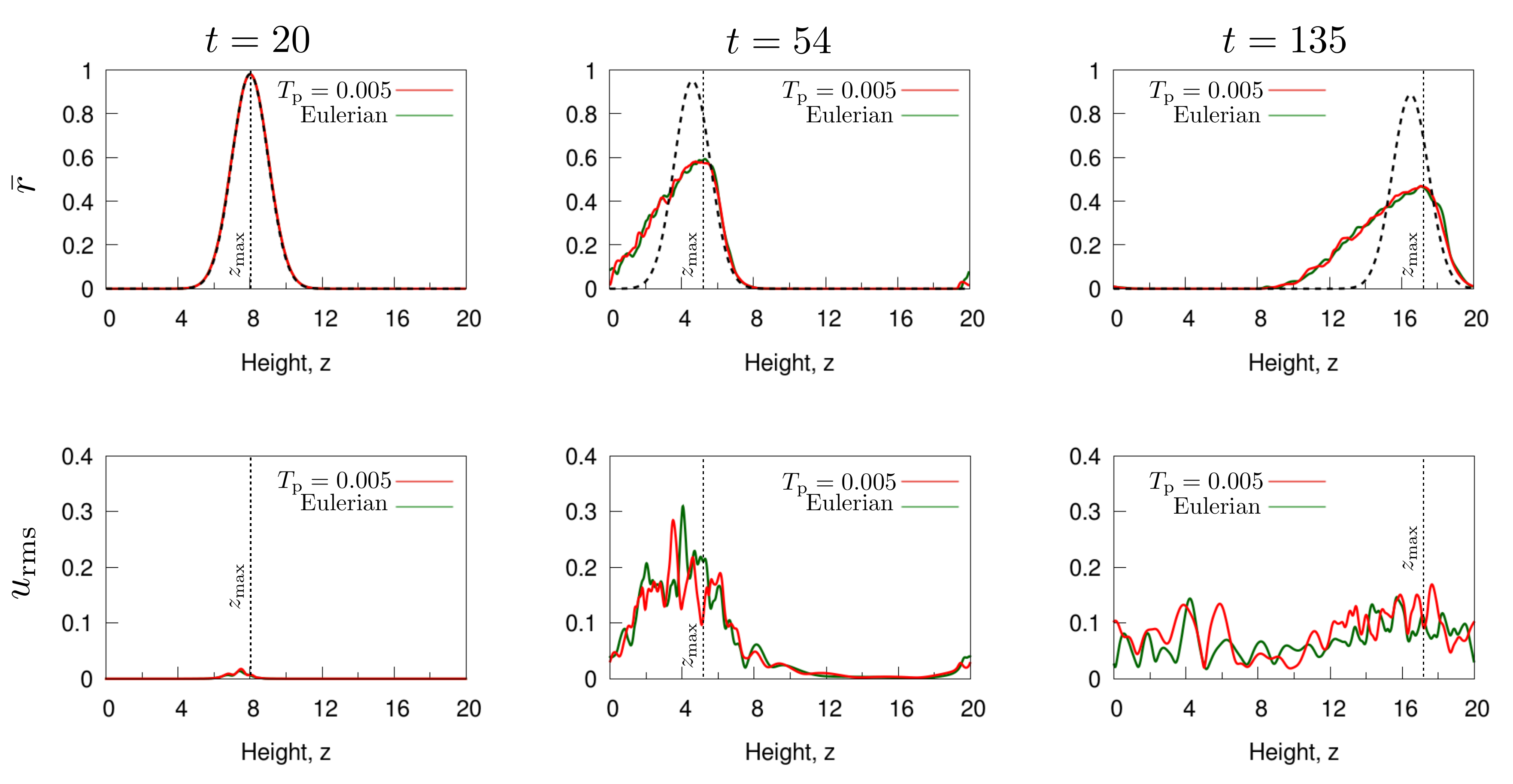}}
\end{figure}
 
Looking at the eddy velocities, we see that the intermediate phase ($t = 30 - 60$) corresponds to the peak of the mixing event. The corresponding $\usup$ reaches a maximum value of $\usup \approx 0.8$ with the rms velocity reaching $\urms^* \approx 0.25$. The fact that $\usup$ and $\urms^*$ are both of order unity actually holds for all runs (see later), and proves that the non-dimensionalization selected is appropriate. By $t = 80$, the main mixing event is over and the turbulence (as measured both by $\urms^*$ or $\usup$) now gradually decays on a much longer  timescale.

We can also look at how the particles and the fluid velocity evolve spatially over time. Figure \ref{sims_evolution} shows the profiles of $\bar{r}$ and $\urms$ at three instants in time for both simulations, with the black dotted curve representing $r_{\rm{diff}}$ \eqref{eqn:diff_sol}. Recall that the domain is periodic in both directions so the particle layer re-emerges at the top after leaving from the bottom. The dotted vertical line $z_{\max}$ marks the position of the maximum of $\bar{r}$. We clearly see that the two-fluid and equilibrium Eulerian simulations  behave in a quantitatively similar way. In both cases, the particle layer settles roughly at the expected rate set by the value of $W_s$, but its vertical density profile $\bar{r}$ becomes asymmetric and wider than in the purely diffusive case (black dotted line). The extended tail of $\bar{r}$ below the bulk of the layer is associated with more rapidly-moving particle-rich plumes that can clearly be seen penetrating into the lower particle-free fluid in Figure \ref{sim2_snapshots}.  Focusing on the evolution of the $\urms$ profile, we see that at early times the turbulence develops in the bulk of the particle layer as expected. However, the fluid remains turbulent even after the particles have settled through a region, which explains why the size of the turbulent region is much larger than that of the particle layer at late times (e.g. $t = 135$). This can be understood by noting that the time it takes for turbulent motions to decay viscously is much larger than the time it takes for the particles to settle across the bottom of the box. 

This section has illustrated the interplay between the turbulence and the particle field for short stopping times. Both the qualitative and quantitative evidence confirm that the two-fluid model for very low $T_{\rm{p}}$ and the equilibrium Eulerian model have similar dynamics, conclusively validating our two-fluid code. 
%---------------------------------------------

\subsection{Comparison between low and high $T_{\rm{p}}$ simulations}\label{subsec:sim_highertp}
\begin{figure} 
\caption{Top: Snapshots of the particle concentration $r$ at various times in a simulation with $\Tp = 0.1$, $W_s = 0.1, R_\rho = 0.5, \Reyn = 1000$, $\Pep = 1000$, ${\rm{Pr}} = 1$. Bottom: Diagnostic properties of the particle concentration and fluid velocity  as a function of time for the same simulation.}
\centering 
\label{sim3}
{\includegraphics[width=0.6\pdfpagewidth]{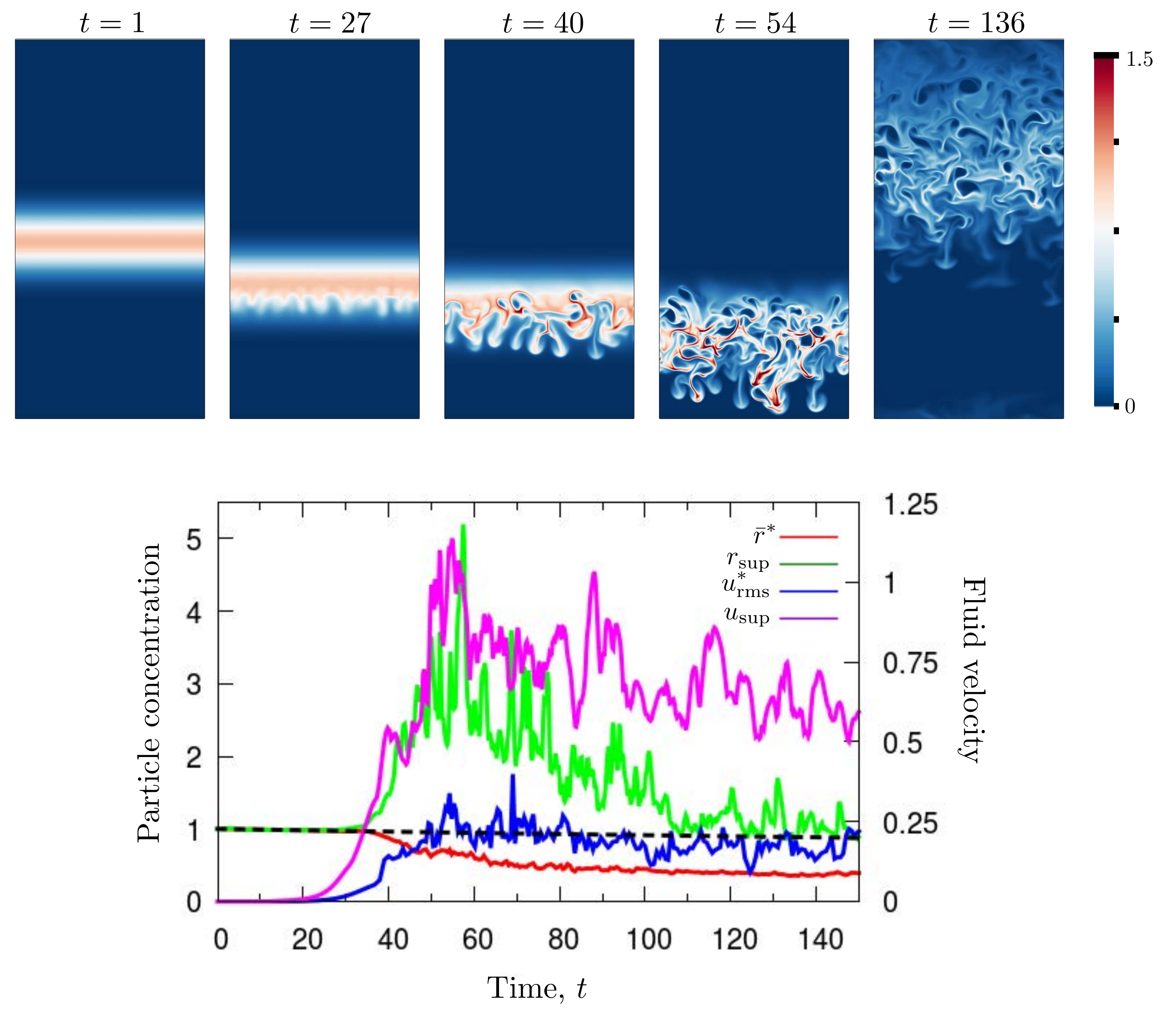}}
\end{figure}

We now look at the effect of larger stopping time on the evolution of the particle layer. We continue to work in 2D and choose $T_{\rm{p}} = 0.1$ with the same resolution (i.e. $768 \times 1536$ grid points) keeping the remaining parameters and domain size the same as in the simulation from Section \ref{subsec:code_validation} (i.e. $W_s = 0.1, R_\rho = 0.5, \Reyn = 1000$, $\Pep = 1000$, ${\rm{Pr}} = 1$; and $L_x = 10$, $L_z = 20$). Snapshots of the particle concentration field as well as the evolution of $\rsup$ and $\bar{r}^*$ with time are shown in Figure \ref{sim3}. We clearly see the emergence of regions of much higher particle concentration than at low $T_{\rm{p}}$, located in narrow, wisp-like structures (see for instance the snapshot at $t=54$) with $\rsup$ reaching values of as high as 5. The fact that this is much larger than the initial maximum value of $r$ in the domain is a distinct signature of preferential concentration, since this only occurs when $\div \boldup$ is non-zero. This also shows that regions of strongly enhanced particle concentration can develop even when the mean particle concentration in the bulk of the layer is decreasing. After the main mixing event (around $t = 80$), $\rsup$ drops again to values that are lower than one, though remains substantially higher than $\bar{r}^*$. This raises the interesting question of what determines the maximum possible value of the particle concentration field at any given time in the simulation (which will be further discussed in Section \ref{sec:analysis}). 

Turning our attention to the evolution of $\usup$ and $\urms^*$, we see that for larger $T_{\rm{p}}$ at the peak of the mixing event, $\usup \approx 1.1$ and $\urms^* \approx 0.3$, whereas in the lower $T_{\rm{p}}$ case the corresponding values were  $\usup \approx 0.8$ and $\urms^* \approx 0.25$. This suggests that $T_{\rm{p}}$ does not have a major effect on the turbulence of the system (at least for the parameters explored). 

We can measure the maximum particle concentration at a given height in the domain using  
\begin{equation}
	\rxmax(z,t) = \max_x r(x,z,t).
\end{equation} 

In addition, we can also measure the typical (rather than the maximum) enhancement over the mean $\bar{r}$ as a function of height using
\begin{equation}\label{eqn:rrms}
	r_{\rm{rms}}(z,t) = \bigg[ \overline{[r(x,z,t) - \bar{r}(z,t)]^2} \bigg]^{1/2}.
\end{equation} 

\begin{figure} 
\caption{Measures of particle concentration at $t = 54$ of two-fluid simulations for $T_{\rm{p}} = 0.005$ and $T_{\rm{p}} = 0.1$ with otherwise identical parameters $W_s = 0.1, R_\rho = 0.5, \Reyn = 1000$, $\Pep = 1000$, and ${\rm{Pr}} = 1$.}
\label{sim_rmethod}
\centering
{\includegraphics[width=0.63\pdfpagewidth]{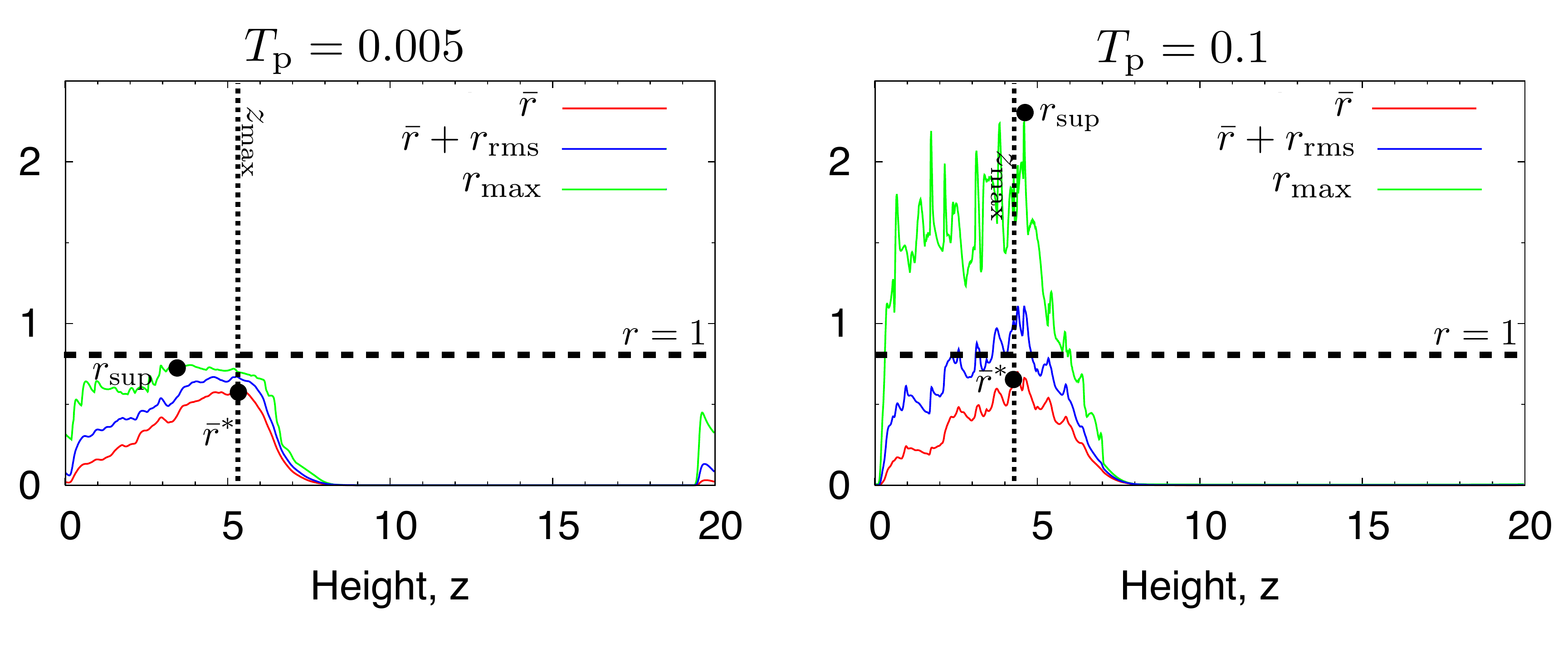}}
\end{figure}

Figure \ref{sim_rmethod} compares the maximum particle concentration $\rmax$ with both the mean particle concentration $\bar{r}$ and one standard deviation above the mean, $\bar{r} + \rrms$, as a function of height, for two simulations with $T_{\rm{p}} = 0.005$ and $T_{\rm{p}} = 0.1$. We see that for both cases, $\bar{r}$ and $\bar{r} + \rrms$ have similar profiles. For the low $T_{\rm{p}}$ case, $\bar{r} + \rrms$ typically remains below one. In addition, the profile of $\rmax$ also follows that of $\bar{r}$, and lies about two standard deviations above it. As such, it is largest in the bulk of the particle layer. For high $T_{\rm{p}}$,  $\rmax$ is also largest in the bulk of the particle layer, with values peaking at $\rsup \approx 2.25$ at this particular instant in time. However, $\rmax$ is now several standard deviations above $\bar{r}$, implying that the probability density distribution of the particle concentration has a longer tail (see Section \ref{sec:pdf} for more on this point). 

\begin{figure} 
\caption{Comparison of $\rsup$ for 2D simulations with varying $T_{\rm{p}}$.  Remaining parameters: $W_s=0.1, R_\rho = 0.5, \Reyn = 1000, \Pep = 1000, {\rm{Pr}} = 1$. An equilibrium simulation marked ``Eulerian" is shown for comparison.}
\centering  
\label{rsupVstime}
\vspace{.1in}
{\includegraphics[width=0.4\pdfpagewidth]{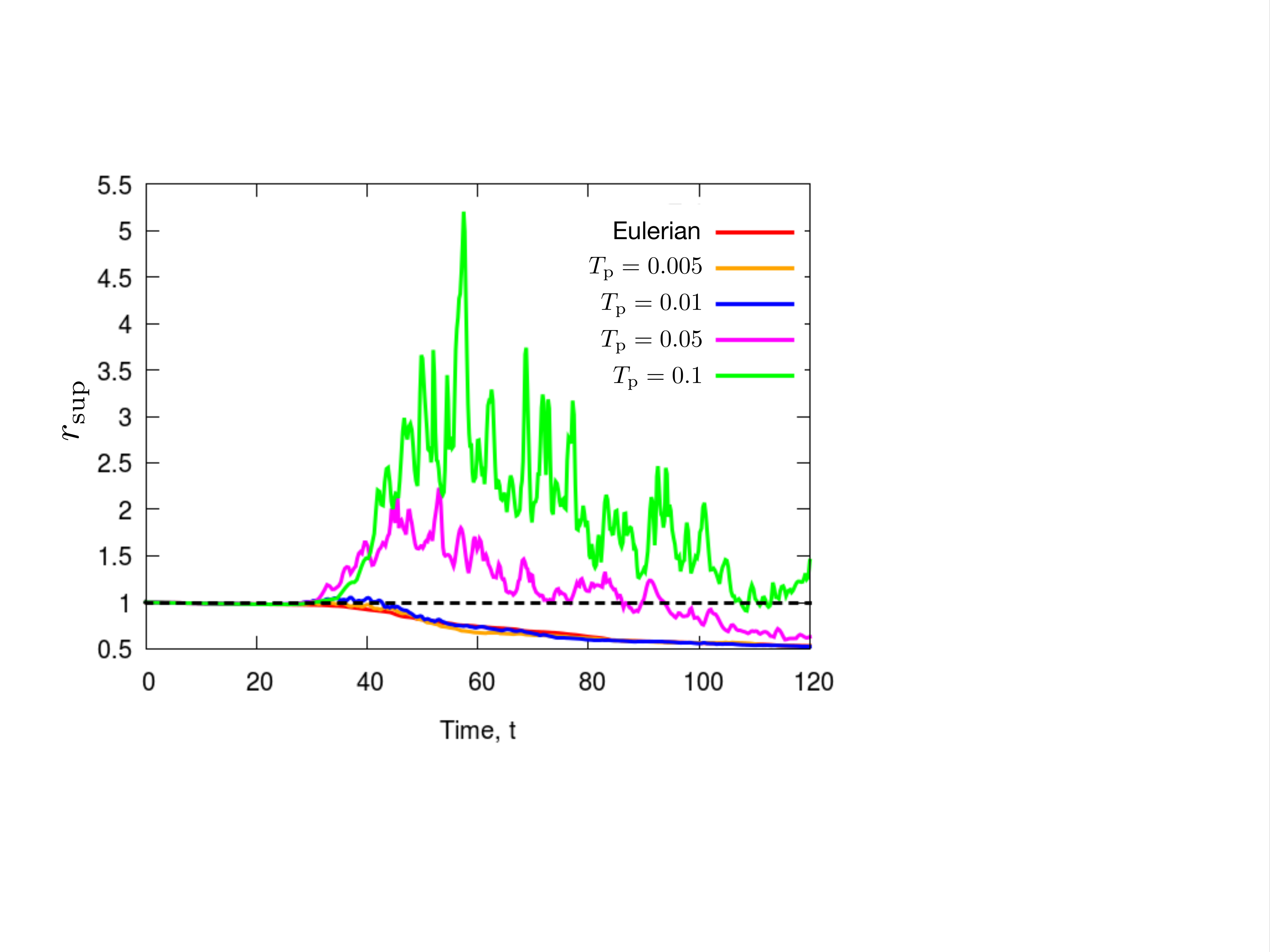}}
\end{figure}

Figure \ref{rsupVstime} more generally compares the maximum particle concentration $\rsup$ obtained in several simulations with increasing particle stopping time $T_{\rm{p}}$.  The simulations continue to be in 2D  with $768 \times 1536$ grid points, and all other parameters remain unchanged (i.e. $W_s = 0.1, R_\rho = 0.5, \Reyn = 1000$, $\Pep = 1000$, ${\rm{Pr}} = 1$). The black dotted line represents $r = 1$. As expected, we find that $\rsup$ increases with $\Tp$ as a result of preferential concentration. Furthermore, we see that $\rsup$ remains above unity for longer times, signifying that dense particle regions persist in the simulations. On the other hand, we find that preferential concentration is negligible for $\Tp \leq 0.01$, and $\rsup$ is almost indistinguishable from that obtained in the equilibrium Eulerian limit.

\subsection{Impact of $\Pep$ and $\Reyn$}\label{subsec:PepRey}
\begin{figure} 
\caption{Snapshots of the particle concentration field for varied $\Reyn$ and $\Pep$ with fixed $T_{\rm{p}} = 0.1$ taken at $t \approx 54$. Only the vicinity of the particle layer is shown. Remaining parameters: $W_s = 0.1, R_\rho = 0.5,$, ${\rm{Pr}} = 1$. }
\centering  
\label{snapshots_varyingRePep}
{\includegraphics[width=0.7\pdfpagewidth]{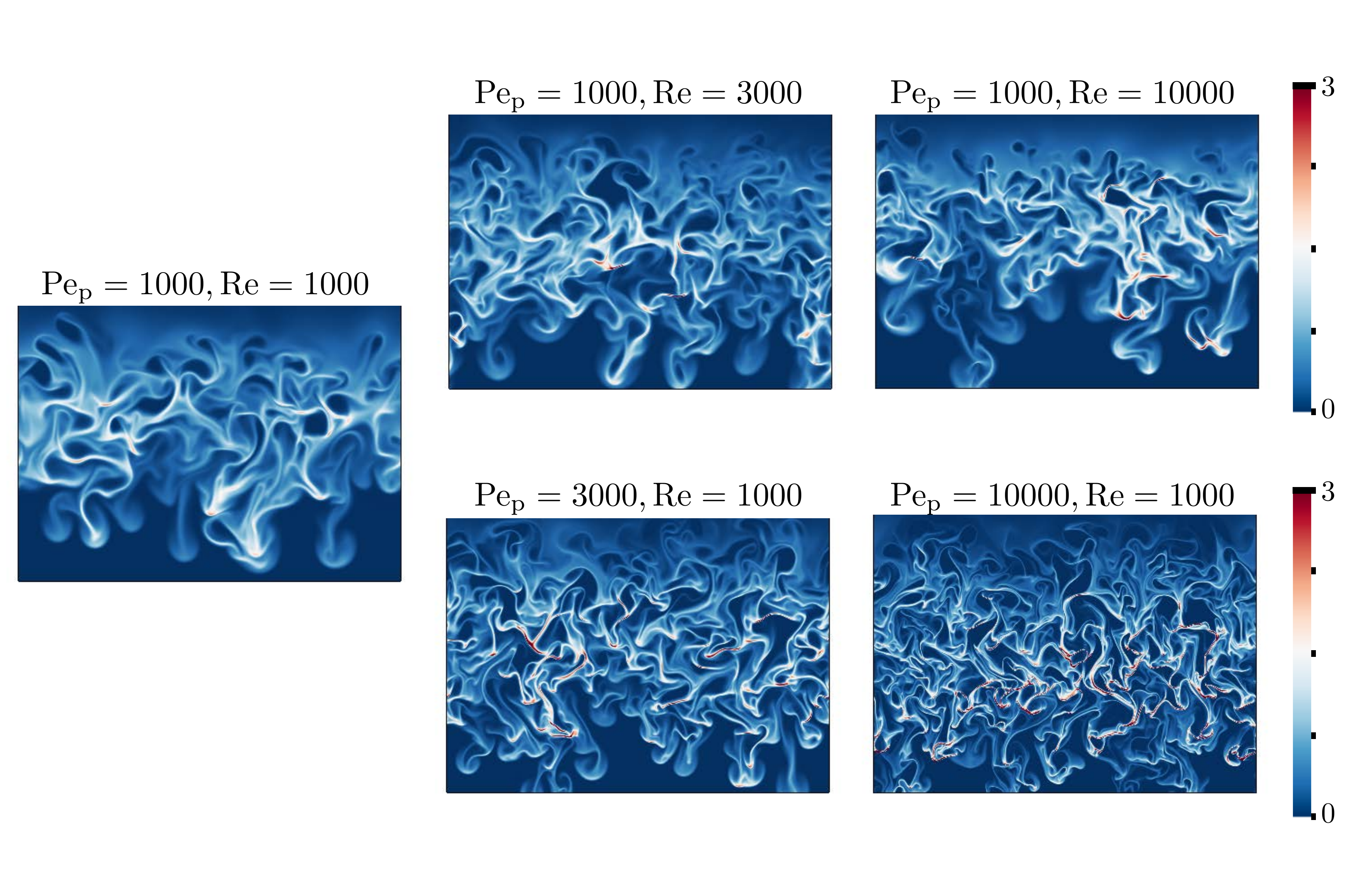}}
\end{figure}

We next look at the impact of the fluid Reynolds number $\Reyn$ and the particle P$\rm{\acute{e}}$clet number $\Pep$ on the evolution of the particle concentration. We continue to focus on 2D simulations, choosing a relatively large stopping time to ensure that inertial effects are important. We use $T_{\rm{p}} = 0.1$ with the remaining parameters and domain size set as $W_s = 0.1, R_\rho = 0.5$, ${\rm{Pr}} = 1$, $L_x = 10$, and $L_z = 20$. The resolution selected for these simulations increases with both $\Reyn$ and $\Pep$, and is listed in Table \ref{table:sims}. Figure \ref{snapshots_varyingRePep} presents snapshots of the particle concentration at $t = 54$ for simulations with $\Pep$ and $\Reyn$ both varying between 1000 and 10,000.  When we fix $\Pep = 1000$ and increase $\Reyn$,  the particle concentration snapshots appear qualitatively similar, consisting of narrow structures comparable in size and density. The maximum particle concentration enhancement appears relatively unaffected by the fluid viscosity (at least, for this range of $\Reyn$, and within the context of the two-fluid equations). In contrast, if we fix $\Reyn = 1000$ and increase $\Pep$, we see a striking difference in both the geometry of the wisps, as well as the maximum concentration achieved in the wisps. That is, as $\Pep$ increases, these structures become more numerous and narrower, with a corresponding increase in the maximum particle concentration.  

\begin{figure} 
\caption{Comparison of $\rsup$ for 2D simulations with varying $\Pep$ and $\Reyn$ for $T_{\rm{p}} = 0.1$. Remaining parameters: $W_s = 0.1, R_\rho = 0.5, \Rep = 1000, {\rm{Pr}} = 1$. }
\label{rsupVaryingPepRe}
\vspace{.1in}
\centering
{\includegraphics[width=0.4\pdfpagewidth]{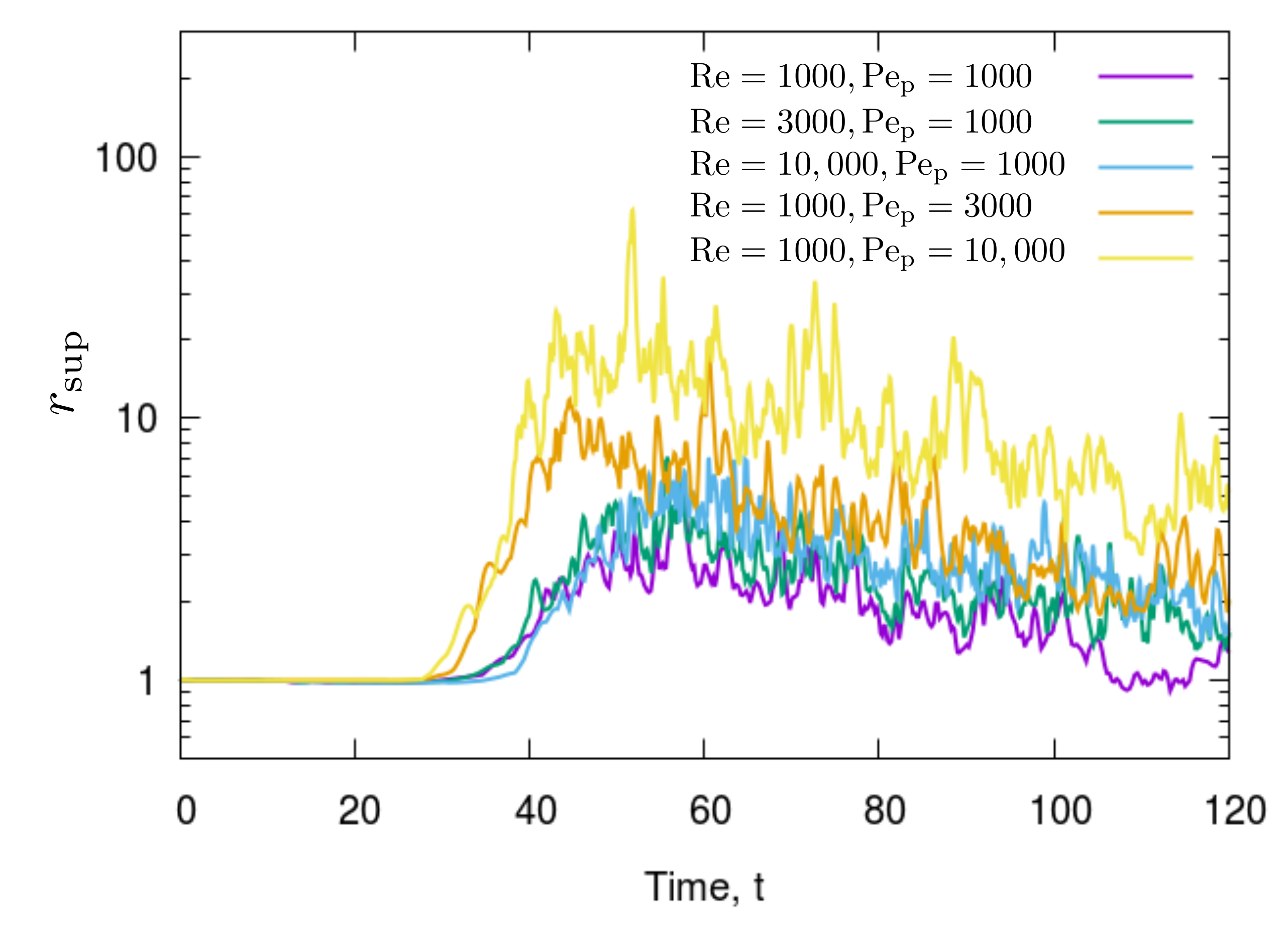}}
\end{figure}

These qualitative trends are confirmed more quantitatively in Figure \ref{rsupVaryingPepRe}, which shows the maximum particle concentration $\rsup$ as a function of time for each of these five simulations. We see that the evolution of $\rsup$ is more or less independent of the Reynolds number but increases with P$\rm{\acute{e}}$clet number. This trend will be explained by the theory presented in Section \ref{sec:analysis}. 

\begin{figure} 
\caption{Power spectra of the particle concentration field (a) and the fluid velocity field (b) as a function of the horizontal wavenumber $k_x$ for varying $\Pep$ and $\Pep$. Remaining parameters are $W_s = 0.1, R_\rho = 0.5, \Rep = 1000, {\rm{Pr}} = 1$.}
\centering
\label{spectra_varyPepRe}
{\includegraphics[width=0.75\pdfpagewidth]{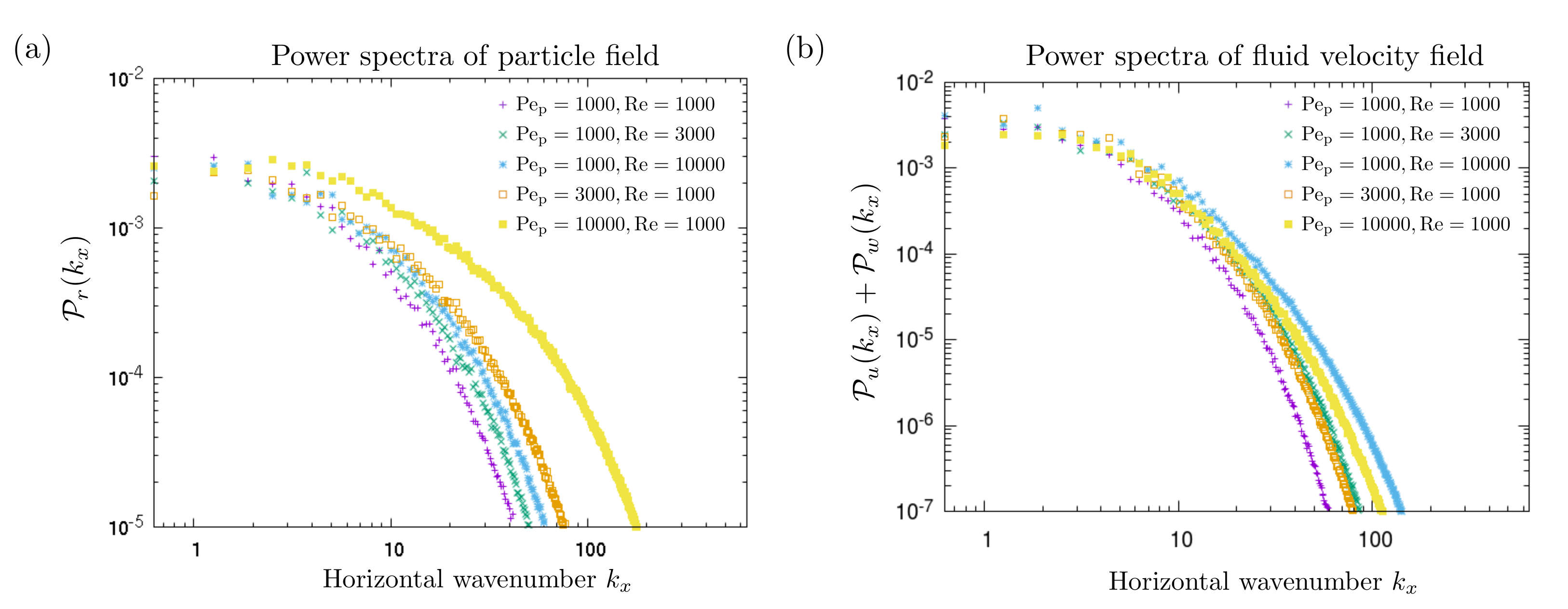}}
\end{figure}

In order to gain a more quantitative insight into the two-way coupling between the particles and the turbulence at all scales, we look at the power spectra of the particle concentration field and of the fluid velocity field. This time we restrict our analysis to an interval of time $[t_0, t_f]$ during the peak of the mixing event when the particle concentration is largest. We define the time-averaged horizontal power spectrum of a quantity $\xi$ as
\begin{align}\label{spec_def}
	\mathcal{P}_{\xi}(k_x) = \frac{1}{t_f - t_0} \int_{t_0}^{t_f} \sum_{k_z} \hat{\xi}(k_x, k_z,t) \hat{\xi}^*(k_x, k_z,t) dt,
\end{align}
where the $\hat{\xi}(k_x, k_z,t)$ is the discrete Fourier transform of $\xi$ and $\hat{\xi}^*(k_x, k_z,t)$ is the complex conjugate of $\hat{\xi}(k_x, k_z,t)$. Figure \ref{spectra_varyPepRe}a shows the horizontal power spectrum of the particle concentration field $\mathcal{P}_{r}(k_x)$. When $\Pep$ is fixed and $\Reyn$ increases, we observe a slight increase of power in the range $k_x = 10-100$, but the effect of $\Reyn$ is small. On the other hand, for fixed $\Reyn$ and large values of $\Pep$, there is substantially more power in the higher wavenumbers, consistent with the predominance of smaller scales seen in the snapshots.

In Figure \ref{spectra_varyPepRe}b, we plot the power spectrum of the total fluid velocity field $\mathcal{P}_{u}(k_x) + \mathcal{P}_{w}(k_x)$ as function of $k_x$. Unlike the particle concentration field, the spectrum here is affected by \textit{both} $\Pep$ and $\Reyn$. That is, the amount of energy at small scales increases when either $\Pep$ or $\Reyn$ increases. This can be explained by the fact that the strength of convection in our system is directly related to the Rayleigh number, which is proportional to the product of $\Pep$ and $\Reyn$ \eqref{eqn:Ra_nondimensional}.  It is therefore not surprising to find that the energy spectrum depends on the product $\Pep \Reyn$ rather than $\Pep$ and $\Reyn$ individually.   
 
\begin{figure} 
\caption{Power spectra of the particle concentration field (left) and the fluid velocity field (right) as a function of the horizontal wavenumber $k_x$ for varying $T_{\rm{p}}$. Remaining parameters are $W_s = 0.1, R_\rho = 0.5, \Reyn =1000, \Rep = 1000, \Pep =1000,$ and ${\rm{Pr}} = 1$.}
\centering
\label{spectra_varyTp} 
{\includegraphics[width=0.75\pdfpagewidth]{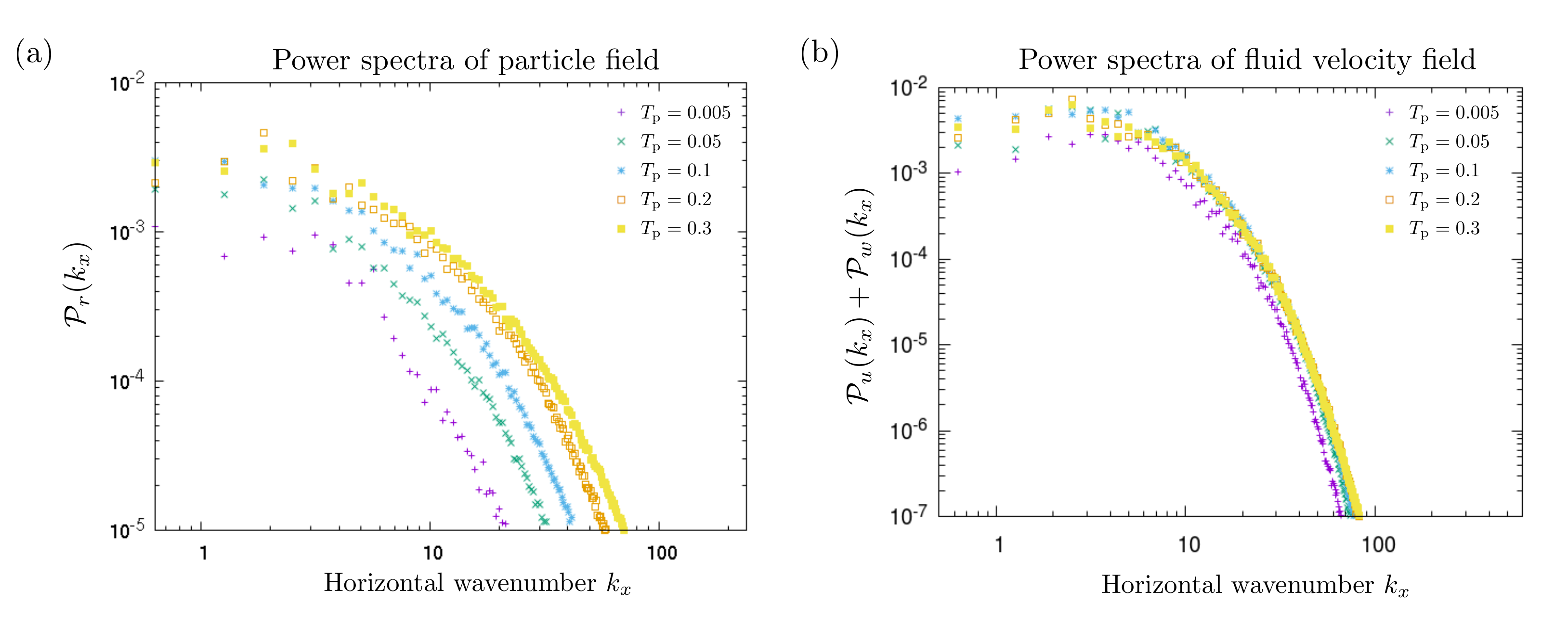}}
\end{figure}

We also look at how the particle stopping time $T_{\rm{p}}$ affects the horizontal power spectra of the particle concentration and velocity fields. Figure \ref{spectra_varyTp} shows these power spectra (taken, as before, during the peak of the mixing event), for five simulations at varying $T_{\rm{p}}$ and otherwise fixed parameters (i.e. $W_s = 0.1, R_\rho = 0.5, \Reyn = 1000$, $\Pep = 1000$, ${\rm{Pr}} = 1$; and $L_x = 10$, $L_z = 20$ with resolution for simulations found in Table \ref{table:sims}). In Figure \ref{spectra_varyTp}a, we see more power at large $k_x$ as $T_{\rm{p}}$ increases. We see this as further evidence that the particles increasingly concentrate in narrower wisps as $\Tp$ increases. In Figure \ref{spectra_varyTp}b, profiles of the total velocity power spectrum, i.e. $\mathcal{P}_{u}(k_x) + \mathcal{P}_{w}(k_x)$, are strikingly similar to one another. Thus, $\Tp$ does not appear to affect the turbulence in the system which is somewhat unexpected given the two-way coupling; instead, the velocity power spectrum is primarily dependent on $\rm{Ra}$, at least for the range of parameters explored here. 

\subsection{Comparison between 2D and 3D simulations} \label{subsec:comp_2D_3D}
Owing to the high resolution needed for the two-fluid simulations, especially for higher $T_{\rm{p}}$, $\Reyn$, and $\Pep$, 3D simulations are typically prohibitive. However, we have run several 3D simulations at moderate $T_{\rm{p}}$, $\Reyn$, and $\Pep$ in order to compare the 3D results with the 2D ones. In this manner, we can determine whether 2D results can at least qualitatively capture the properties of the particle layer evolution.  For all 3D simulations, we set the non-dimensional length, width, and height as $L_x = 10$, $L_y = 2$, and $L_z = 10$, respectively. In this section, we focus on two simulations with $\Tp = 0.005$ and $\Tp = 0.1$, respectively. The resolution of the low $T_{\rm{p}}$ case is $384 \times 72 \times 384$ grid points, while the high $T_{\rm{p}}$ case has a resolution of  $768 \times 144 \times 768$ grid points. 

\begin{figure} 
\caption{Comparison of the fluid velocity $\usup$ and particle concentration $\rsup$ (defined in text) between 2D and 3D simulations with settling velocity $W_s = 0.1, R_\rho = 0.5, \Reyn =1000, \Rep = 1000, \Pep =1000,$ and ${\rm{Pr}} = 1$. Left figure: $T_{\rm{p}} = 0.005$. Right figure: $T_{\rm{p}} = 0.1$.}
\vspace{.2in}  
\label{sim_2Dvs3D}
\centerline
{\includegraphics[width=0.7\pdfpagewidth]{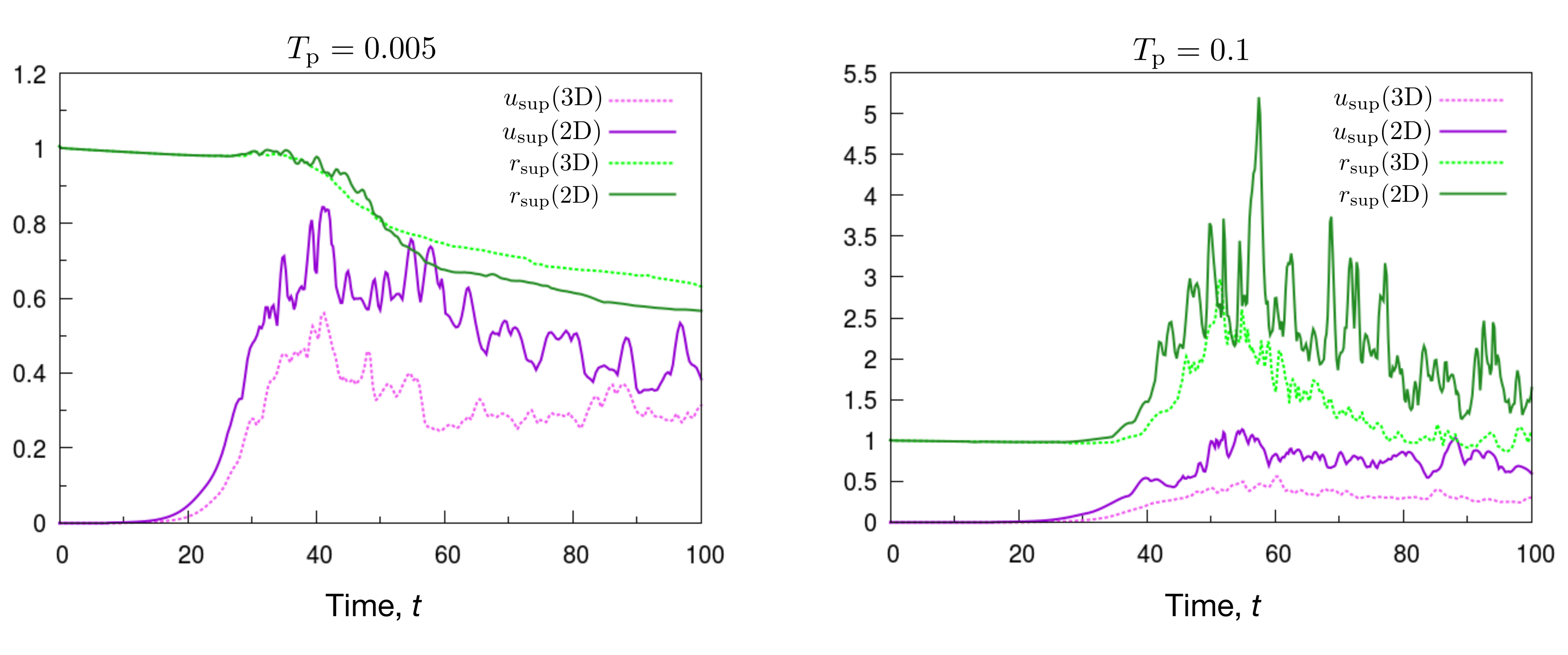}}
\end{figure}

Figure \ref{sim_2Dvs3D} shows that the values of $\usup$  achieved in the 2D simulation are consistently larger than the 3D simulation by 30-50\% (for both low and high $T_{\rm{p}}$ cases). This result is consistent with those of \citep{van2013comparison} for Rayleigh-B\'{e}nard convection (where the rms velocities in 2D are systematically larger than in 3D by a factor of about 2). As a result, turbulent mixing and preferential concentration are both more energetic in 2D than in 3D at otherwise similar parameters. For low $\Tp$ where preferential concentration is not present, enhanced turbulent mixing results in $\rsup$ being slightly smaller in 2D than in 3D. By contrast at high $\Tp$, $\rsup$ is slightly larger in 2D than in 3D due to the enhanced preferential concentration. Generally speaking, however, the dimensionality of the model does not appear to affect preferential concentration by more than a constant factor of a few (see more on this below), suggesting that 2D simulations are appropriate, at least as far as extracting scaling laws is concerned.  

\section{Predicting maximum particle concentration}\label{sec:analysis}
We now present a simple model to quantify the effects of preferential concentration in convective particle-driven instabilities. We begin with the particle concentration equation \eqref{eqn:nondim3}, substituting $r = \bar{r}(z,t) + r'(\mathbf{x},t)$ (where $\mathbf{x}$ is the position vector): 
\begin{equation}
	\frac{\partial (\bar{r} + r')}{\partial t} + (\bar{r} + r') \div \boldup + \boldup \cdot \grad (\bar{r} + r') = \frac{1}{\Pep} \grad ^2 (\bar{r} + r').
\end{equation}
By expanding the divergence term, we note that only the second term on the left-hand side contributes to preferential concentration (when $\div \boldup \neq 0$). We next assume that in the fully turbulent high $T_{\rm{p}}$ flow, regions of particularly strong particle concentration enhancement are characterized by a dominant balance between the preferential concentration of the mean particle density and diffusion terms of the perturbations so that
\begin{equation}\label{eqn:dominant}
	\bar{r}  \div \boldup \sim \frac{1}{\Pep} \grad ^2 r'.
\end{equation}
We then express the particle velocity $\boldup$ in terms of $T_{\rm{p}}$ and $\boldu$, using a standard asymptotic expansion in $T_{\rm{p}}$ \citep{maxey1987gravitational}:
\begin{equation}\label{eqn:nondimup}
	\boldup = \boldu - W_s \ez - T_{\rm{p}} \bigg(\boldu \cdot \grad \boldu + \frac{\partial \boldu }{\partial t} \bigg) + O(T_{\rm{p}} ^2),
\end{equation}
and thus, 
\begin{equation}\label{eqn:divup} 
	\nabla \cdot \boldup = - T_{\rm{p}} \nabla \cdot ( \boldu \cdot \nabla \boldu) + O(T_{\rm{p}} ^2).
\end{equation}
Substituting \eqref{eqn:divup} in \eqref{eqn:dominant} results in
\begin{equation}\label{eqn:almostparticlemodel}
	\bar{r} \div (\boldu \cdot \grad \boldu) T_{\rm{p}} \sim \frac{1}{\Pep} \grad ^2 r'.
\end{equation}
Assuming that the length scales of the inertial concentration and diffusion terms are the same, we finally get 
\begin{equation} \label{eqn:particlemodel}
	\frac{r'}{\bar{r}} \sim {|\boldu| ^2 T_{\rm{p}} \Pep} \sim \frac{\urms^2 \tau_p}{\kappa_p},
\end{equation}
where the third part of this equation is expressed dimensionally. In this model, we therefore predict that strong particle concentration enhancements above the mean only depend on the magnitude of the fluid velocity $\boldu$, the particle stopping time $T_{\rm{p}}$, and the assumed particle diffusion coefficient $\Pep$. The prediction \eqref{eqn:particlemodel} made for $r'/\bar{r}$ should hold in a large-scale sense (i.e. a scale greater than several eddy scales), and can help quantify the expected spatiotemporal evolution of $r'$ as long as that of $\bar{r}$ and $|\boldu|$ is known. 

In order to test our model, we have run a large number of 2D simulations (with a few 3D ones) at different values of $W_s$, $T_{\rm{p}}$, and $\Pep$, listed in Table \ref{table:sims}. Since the particle layer is not much wider than the size of an eddy, we investigate the validity of the model here only as a function of time, focusing on the behavior within the bulk of the particle layer (i.e. near $z = z_{\max}$). To estimate the maximum particle concentration enhancement in the bulk of the particle layer, we let $r'(z,t) = \rmax(z,t) - \bar{r}(z,t)$ and find the maximum value of $r'/\bar{r}$ at each instant in time to obtain
\begin{equation}
	 \bigg(\frac{r'}{\bar{r}}\bigg) _{\max} \simeq \max_{z \in [\zmax - 1, \zmax + 1]} \frac{ r_{\rm max}(z,t) - \bar{r}(z,t)}{\bar{r}(z,t)}. 
\end{equation}
To estimate the corresponding typical fluid velocity, we define the rms total fluid velocity found within the particle layer, defined as
\begin{equation}
	\urmstot(t) = \bigg( \frac{1}{2 L_x L_y} \int^{\zmax+1}_{\zmax-1} \int_0^{L_y} \int_0^{L_x} [u^2(\mathbf{x},t) + v^2(\mathbf{x},t) + w^2(\mathbf{x},t)] dx dy dz \bigg) ^ {1/2},
\end{equation}
where $L_y = 1$ and $ v(\mathbf{x},t) = 0$ for  2D simulations.

\begin{figure} 
\caption{Maximum particle concentration enhancement over the mean as function of $\urmstot^2 \Tp \Pep$ for a simulation with parameters $\Tp = 0.3, W_s = 0.1, R_\rho=0.5, {\rm{Pr}} = 1,  \Pep = 1000,$ and $\Reyn = 1000$. Each dot represents an instant in time, with points moving from the bottom-left corner to the top-right corner over time. The black solid line shows $(r'/\bar{r})_{\max} = (1/4)\urmstot^2\Tp\Pep$.}
\vspace{.2in}
\hspace{-.5in}
\label{concen_rmax_onesim} 
\centering
{\includegraphics[width=0.45\pdfpagewidth]{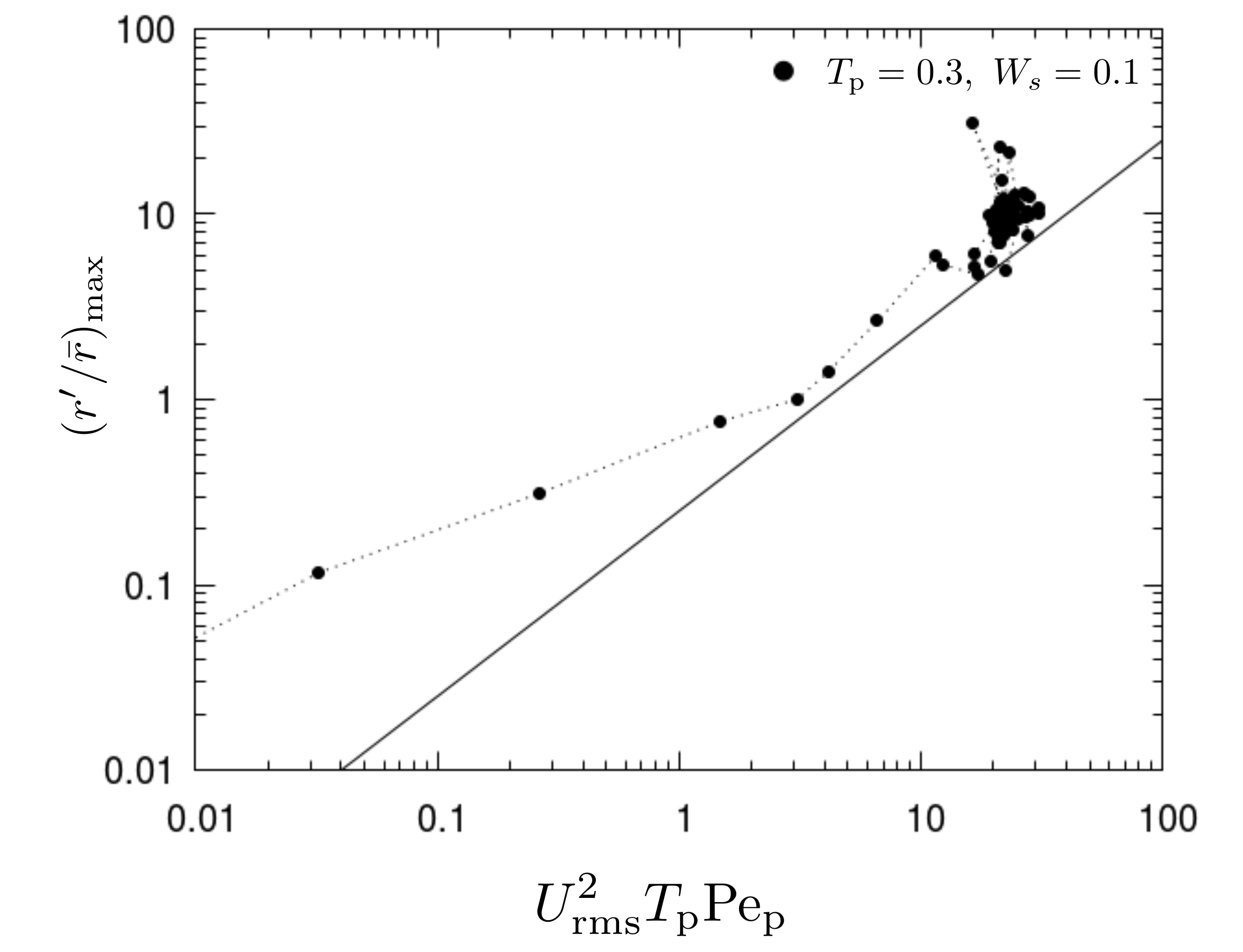}}
\end{figure}

In Figure \ref{concen_rmax_onesim}, we plot $(r'/\bar{r})_{\max}$ versus $\urmstot^2 T_{\rm{p}} \Pep$ for one simulation ($T_{\rm{p}} = 0.3$, $W_s = 0.1$, $\Pep = 1000$, and $\Reyn = 1000$). Note that each data point represents an instant in time for which the full velocity and particle fields are available. Points start from the lower left corner and move up to the right as $\urmstot$ increases with time during the development of the convective instability.  During the most turbulent stage of the simulation when particle concentration enhancement occurs, the points are clustered on the upper right-hand side of the plot. The dashed line represents the scaling relationship$(r'/\bar{r})_{\max} \propto \urmstot^2 T_{\rm{p}} \Pep$, shown here for ease of comparison with later figures. 

\begin{table}
\caption{Numerical simulations. The first four columns show $W_s$, $T_{\rm{p}}$, $\Pep$, and $\Reyn$. The last column shows the effective number of mesh points in each direction. All 2D simulations in this table were run with $L_x = 10$ and $L_z = 20$, and all 3D simulations were run with $L_x = 10$, $L_y = 2$, and $L_z = 10$.}
\label{table:sims}
\centering
\begin{tabular}{@{\hskip .2in}l@{\hskip .2in}l@{\hskip .2in}l@{\hskip .2in}l@{\hskip .2in}c@{\hskip .2in}c@{\hskip .2in}}
\multicolumn{5}{@{}l}{\em(a) 2D Simulations}\\
\toprule
 $W_s$   & $T_{\rm{p}}$ & $\Pep$ & $\Reyn$  & $N_x \times N_z$ \\
\midrule
					 0.1			&	 0.005	& 1000 	& 1000	&  $768 \times 1536$ \\
                    0.1			&	 0.005	& 10,000 	& 1000 &  $768 \times 1536$ \\
                    0.1			&	 0.005	& 100,000 	& 1000 &   $3072 \times 6144$ \\
                    0.1			&	 0.01   	& 1000 	&  1000 &  $768 \times 1536$ \\
                    0.1			&	 0.05   	& 1000 	&  1000 &  $768 \times 1536$ \\
                    0.1			&	 0.1     	& 1000 	&  1000 &  $768 \times 1536$ \\
                    0.1			&	 0.1     	& 3000 	&  1000 &  $1536 \times 3072$ \\
                    0.1			&	 0.1     	& 10,000 &	1000 &  $3072 \times 6144$ \\
                    0.1			&	 0.1     	& 1000 	&  3000 &  $1536 \times 3072$ \\
                    0.1			&	 0.1     	& 1000 &	10,000 &  $3072 \times 6144$ \\
                    0.1			&	 0.2     	& 1000 	&  1000 &  $1536 \times 3072$ \\
                    0.1			&	 0.3     	& 1000 	&  1000 &  $1536 \times 3072$ \\
                   \hline 
                    0.3			&	 0.005	& 1000 	&  1000 &  $768 \times 1536$ \\
                    0.3			&	 0.01	& 1000 	&  1000 &  $768 \times 1536$ \\
                    0.3			&	 0.05	& 1000 	&  1000 &  $768 \times 1536$ \\
                    0.3			&	 0.1   	& 1000 	&  1000 &  $1152 \times 2304$ \\
                    0.3			&	 0.2   	& 1000 	&  1000 &  $1536 \times 3072$ \\
                    0.3			&	 0.3     	& 1000 	&  1000 &  $1536 \times 3072$ \\
\bottomrule  
\end{tabular}

\bigskip\bigskip
\begin{tabular}{@{\hskip .2in}l@{\hskip .2in}l@{\hskip .2in}l@{\hskip .2in}l@{\hskip .2in}c@{\hskip .2in}c@{\hskip .2in}}
\multicolumn{5}{@{}l}{\em(b) 3D Simulations}\\
\toprule
 $W_s$   & $T_{\rm{p}}$ & $\Pep$ & $\Reyn$ &  $N_x \times N_y \times N_z$ \\
\midrule
 0.1			&	 0.005	& 1000 	&  1000 &  $384 \times 72 \times 384$ \\
                    0.1			&	 0.1			& 1000	&  1000 & $768 \times 144 \times 768$ \\
                    0.1			&	 0.2			& 1000 	&  1000 &  $768 \times 144 \times 768$ \\
\bottomrule \\
\end{tabular}
\end{table}

\begin{figure} 
\caption{Maximum particle concentration enhancement over the mean as function of $\urmstot^2 \Tp \Pep$, for varying $W_s$ and $T_{\rm{p}}$ (with $R_\rho=0.5, {\rm{Pr}} = 1,  \Pep = 1000, \Reyn = 1000, \Rep = 1000$). The black solid line represents $(r'/\bar{r})_{\max} = (1/4)\urmstot^2\Tp\Pep$. Details of simulations can be found in Table \ref{table:sims}.}
\centering  
\label{concentration_model} 
\vspace{.3cm}
{\includegraphics[width=0.6\pdfpagewidth]{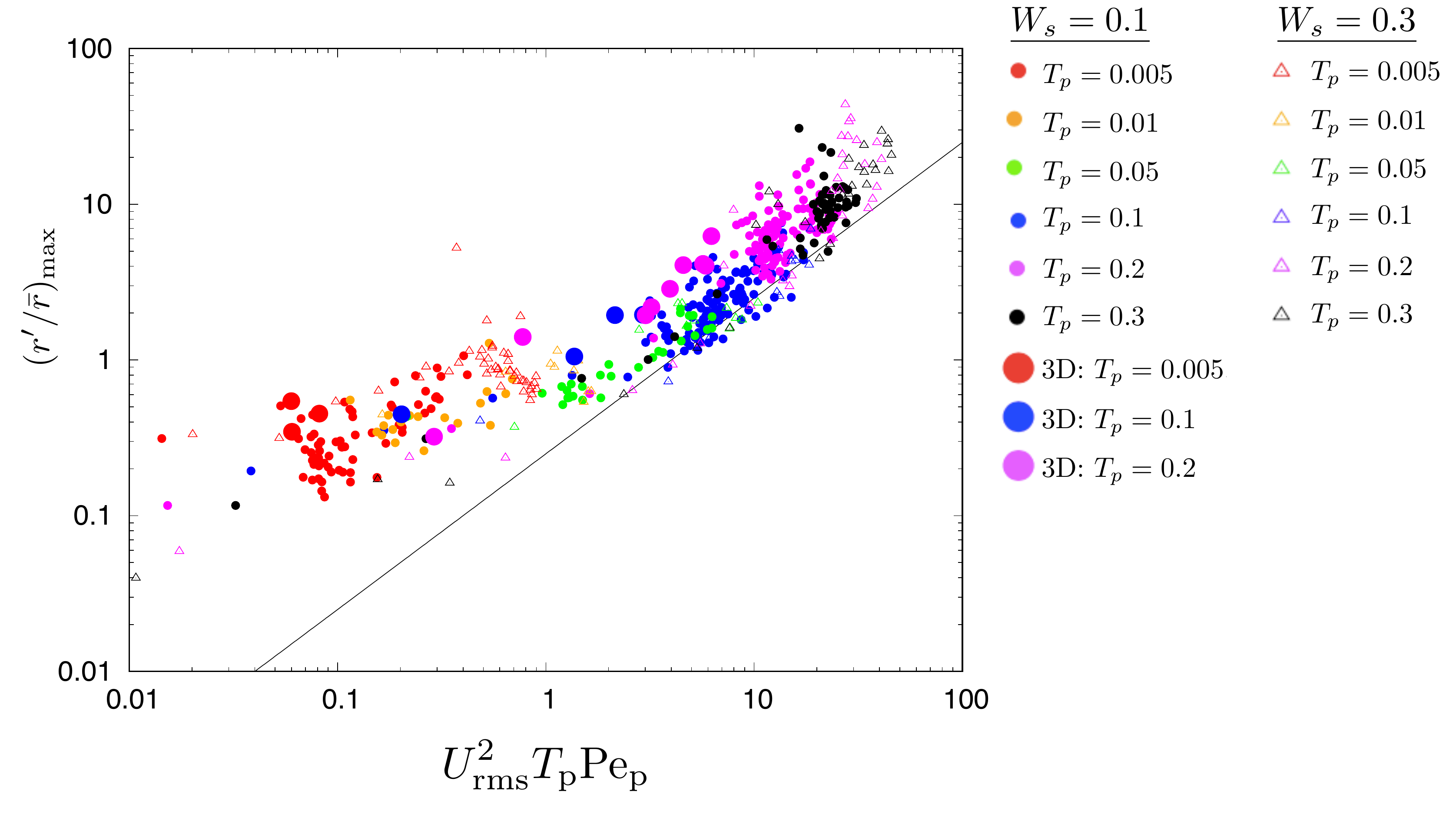}}
\end{figure}

Comparisons between $(r'/\bar{r})_{\max}$ and $\urmstot^2 T_{\rm{p}} \Pep$ are next shown in Figure \ref{concentration_model} for all available simulations that have $\Reyn = 1000$, and $\Pep = 1000$. Here, the color of the points represents $T_{\rm{p}}$, the shape of the points represent $W_s$, and the size of the points corresponds to the dimensionality (2D vs. 3D), see legend for detail. For a given simulation, each point corresponds to a particular instant in time selected after the onset of the convective instability, but before the bulk of the particle layer has traveled more than one domain height (to avoid it interacting with itself). The solid line shows the relationship $(r'/\bar{r})_{\max} = (1/4) \urmstot^2 T_{\rm{p}} \Pep$, where the proportionality constant $1/4$ was selected to fit (approximately) the 2D data in the higher $T_{\rm{p}}$ runs.

Focusing our attention first on the low $T_{\rm{p}}$ 2D simulations (shown in red and orange), we see that they do not fit the model, regardless of the values of $W_s$. This is as expected, since we have found that preferential concentration is negligible for $T_{\rm{p}} \leq 0.01$ (e.g. Figure \ref{rsupVaryingPepRe}), and so the dominant balance assumed in deriving the model in equation \eqref{eqn:particlemodel} does not apply. Turning to the remaining 2D simulations, we see the data fits the predicted model well albeit with a significant scatter that is expected given the method we are using to extract $r'$ and $\urmstot$. We also see that even for cases with larger $\Tp$, there appears to be a threshold (namely $\urmstot^2\Tp\Pep \approx 1$) below which the model is not valid. Above that threshold, the scaling law proposed correctly predicts how $(r'/\bar{r})_{\max}$ evolves in a simulation as a function of time. Finally, we have run several 3D simulations represented by the larger filled circles, and see that they also fit the model. We therefore conclude that equation \eqref{eqn:particlemodel} provides a reliable method for estimating the maximum possible particle concentration enhancement over the mean in a turbulent fluid (within the two-fluid formalism).

\begin{figure} 
\caption{Maximum particle concentration enhancement over the mean as function of $\urmstot^2 \Tp \Pep$, for varying $T_{\rm{p}}$, $\Reyn$, and $\Pep$ (with $W_s = 0.1, R_\rho=0.5, {\rm{Pr}} = 1, \Rep = 1000$). The black solid line represents $(r'/\bar{r})_{\max} = (1/4)\urmstot^2\Tp\Pep$. Details of simulations can be found in Table \ref{table:sims}.}
\centering  
\label{concentration_model_varyingPepRe} 
\vspace{.3cm}
{\includegraphics[width=0.6\pdfpagewidth]{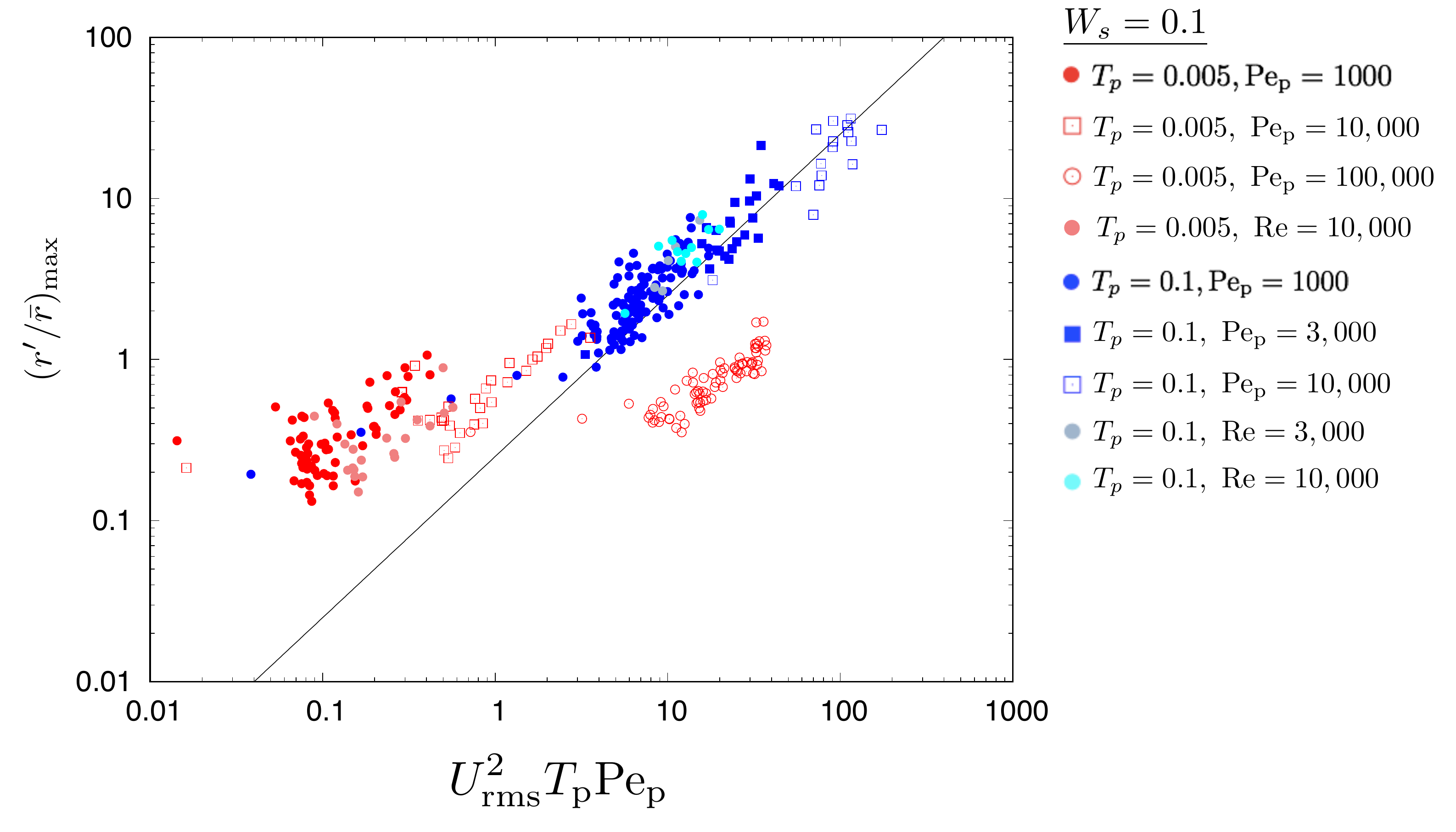}}
\end{figure}

Figure \ref{concentration_model_varyingPepRe} explores the dependence of the model on $\Reyn$ and $\Pep$. As before, the low $T_{\rm{p}}$  simulations (in red) do not fit the model while those at higher $\Tp$ (all other colors) do. We also see that, as discussed in Section \ref{subsec:PepRey}, $(r'/\bar{r})_{\max}$ is more or less independent of $\Reyn$, but increases with $\Pep$.  

\section{Typical particle concentration and pdfs of the relative particle concentration field}\label{sec:pdf}
\begin{figure} 
\caption{Typical particle concentration enhancement for varying $W_s$ and $\Tp$ with $R_\rho=0.5, {\rm{Pr}} = 1,  \Reyn =1000, \Rep=1000, \Pep = 1000$, unless otherwise denoted. The black solid line represents $(r'/\bar{r})_{\rm{rms}} = (1/4) \urmstot^2\Tp\Pep$, and the blue line represents $(r'/\bar{r})_{\rm{rms}} = (1/5)( [\urmstot^2\Tp\Pep]^{1/2}$. Details of simulations can be found in Table \ref{table:sims}.}
\centering  
\label{concentration_model_rrms} 
\vspace{.3cm}
{\includegraphics[width=0.6\pdfpagewidth]{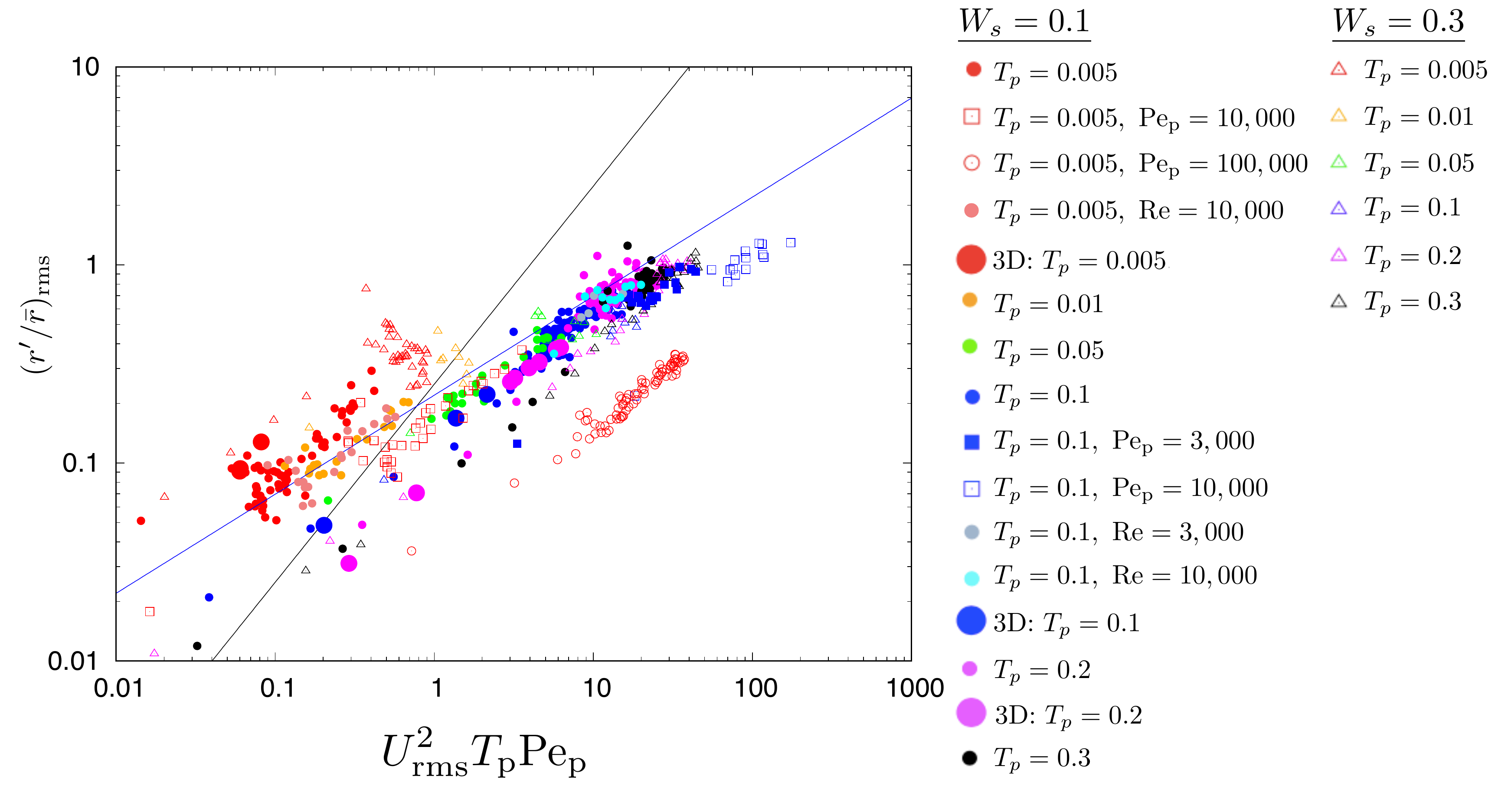}}
\end{figure}

Having constructed a simple analytical model for the maximum particle concentration enhancement allowable in the system, we may wonder whether this model might also provide insight into the typical concentration enhancement. To do so, we define the typical concentration enhancement within the particle layer as:
%To do so, we replace our estimate for $r'$ in equation \eqref{eqn:particlemodel} with $\rrms^*$, defined as  
%\begin{equation}
  % \rrms^*(t) = \rrms(z_{\max}, t).
%\end{equation}
\begin{equation}
	\bigg(\frac{r'}{\bar{r}} \bigg) _{\rm{rms}} = \frac{1}{2} \int_{\zmax - 1} ^{\zmax + 1} \frac{\rrms(z,t)}{\bar{r}(z,t)} dz,
\end{equation}
where $\rrms$ was defined in \eqref{eqn:rrms}. Results are shown in Figure \ref{concentration_model_rrms}, with the same black line as in Figure \ref{concentration_model} also plotted to ease the comparison. Here, we see the data points do not fit this model, and seem to scale as $(r'/\bar{r})_{\rm{rms}} \sim ( \urmstot^2 T_{\rm{p}} \Pep )^{1/2}$ instead (shown by the blue line). It is interesting to note that although we are capturing the \textit{typical} enhancement, this model still depends on the same combination of parameters (i.e. the product of $\urmstot, T_{\rm{p}}$, and $\Pep$) arising from the model discussed in Section \ref{sec:pdf}. This strongly suggests that the typical particle concentration enhancement is related to the maximum particle concentration enhancement, though exactly how remains to be determined. We also see here that the low $T_{\rm{p}}$ simulations (in red and orange) do not follow the same scaling law as the high $T_{\rm{p}}$ cases, but instead, have $(r'/\bar{r})_{\rm{rms}}$ almost constant.

\begin{figure} 
\caption{Probability distribution functions for the function $r_{\rm{rel}}$ \eqref{eqn:rrel} at various times during  two simulations with $W_s=0.1, R_\rho=0.5, {\rm{Pr}} = 1,  \Reyn =1000, \Rep=1000,$ and  $\Pep = 1000$ for $T_{\rm{p}} = 0.005$ (a) and $T_{\rm{p}} = 0.1$ (b).}
\label{pdf_compare}
\vspace{.3cm}
\centering
{\includegraphics[width=0.75\pdfpagewidth]{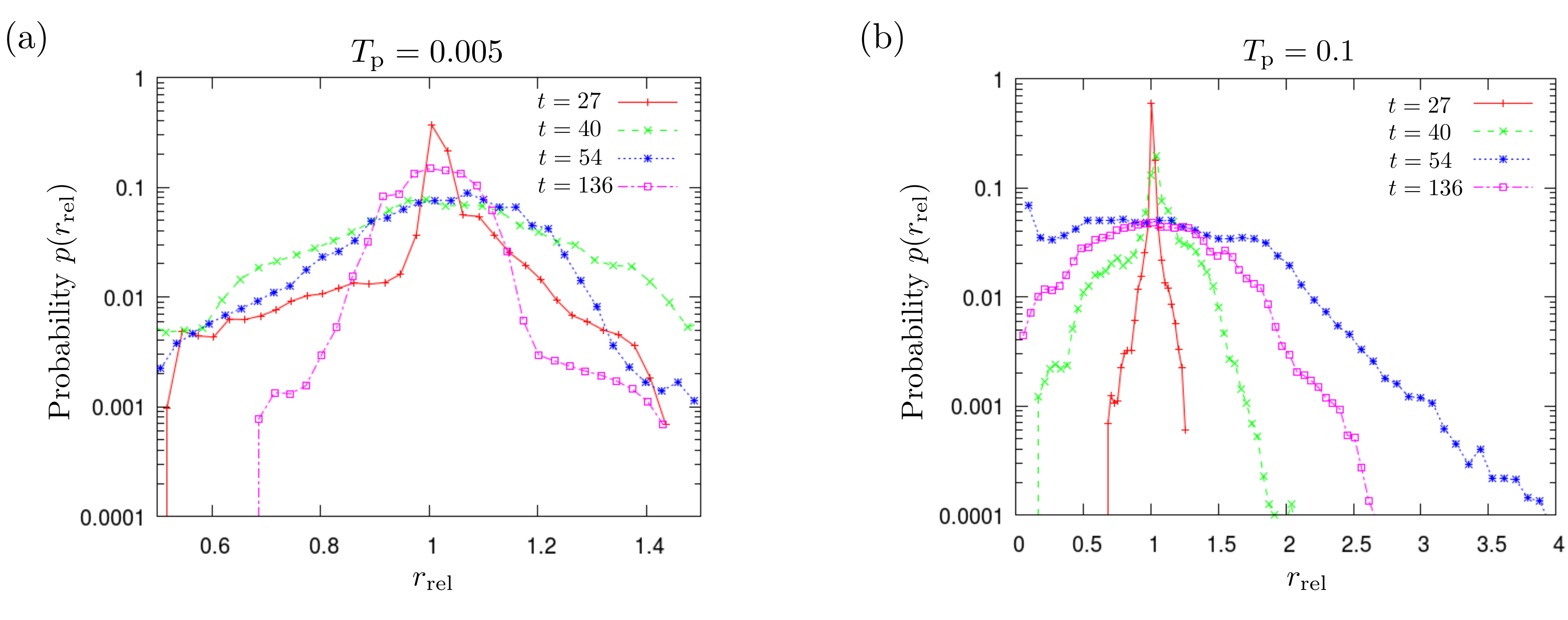}}
\end{figure}

More insight into the problem can be gained by looking at the probability distribution function (PDF) of the relative particle concentration: 
\begin{equation}\label{eqn:rrel}
	r_{\rm{rel}}(x, z, t) = { \frac{r(x,z,t)}{\bar{r}(z,t)}  }  = 1 + \frac{r'(x,z,t)}{\bar{r}(z,t)}.
\end{equation}
We focus on values of $r_{\rm{rel}}$ within the bulk of the particle layer in the range  $z \in [z_{\max}(t) - 1$, $z_{\max}(t) + 1]$. Figure \ref{pdf_compare} shows PDFs of  $r_{\rm{rel}}$ for the low and high $T_{\rm{p}}$ cases presented in Section \ref{subsec:comp_2D_3D} at various times during the respective simulations. Prior to the onset of turbulence the PDF of $r_{\rm{rel}}$ is a $\delta$ function centered at $r_{\rm{rel}} = 1$ since $r = \bar{r}$ . The distribution then widens once the instability develops, and the maximum value achievable by $r_{\rm{rel}}$ is equal to the value $(r'/\bar{r})_{\max} + 1$ discussed in Section \ref{sec:pdf}. 

For the low $T_{\rm{p}}$ case, we see from Figure \ref{pdf_compare}a that the PDF is more or less symmetric about $r_{\rm{rel}}= 1$ at all times, and remains relatively narrow around this mean value (at least, compared with the high $T_{\rm{p}}$ case described below). As the simulation proceeds, the width of the PDF first increases and then decreases with time, as a result of the concurrent increase and decrease of the turbulent fluid velocity $\urms^*$ \eqref{eqn:urmsstar} in the bulk of the layer during the convective mixing event.  In contrast, for the high $T_{\rm{p}}$ simulation shown by Figure \ref{pdf_compare}b, the PDF widens considerably during the convective mixing event and becomes asymmetric. A long tail of rare events associated with preferential concentration appears. The shape of the tail appears to be exponential, consistent with what is commonly found in Eulerian-Lagrangian simulations of preferential concentration (e.g. \citep{shotorban2006particle, zaichik2005statistical}). 

\begin{figure} 
\caption{Time-averaged PDFs of $r_{\rm{rel}}$ (see Equation \ref{eqn:rrel}) during the peak of the mixing event. (a) Fixed $\Reyn = 1000$ and $\Pep = 1000$ and varying $\Tp$. (b) Fixed $\Tp = 0.1$ and varying $\Reyn$ and $\Pep$.}
\centering  
\label{pdf_compare_Pep_Tp}
\vspace{.3cm}
{\includegraphics[width=0.75\pdfpagewidth]{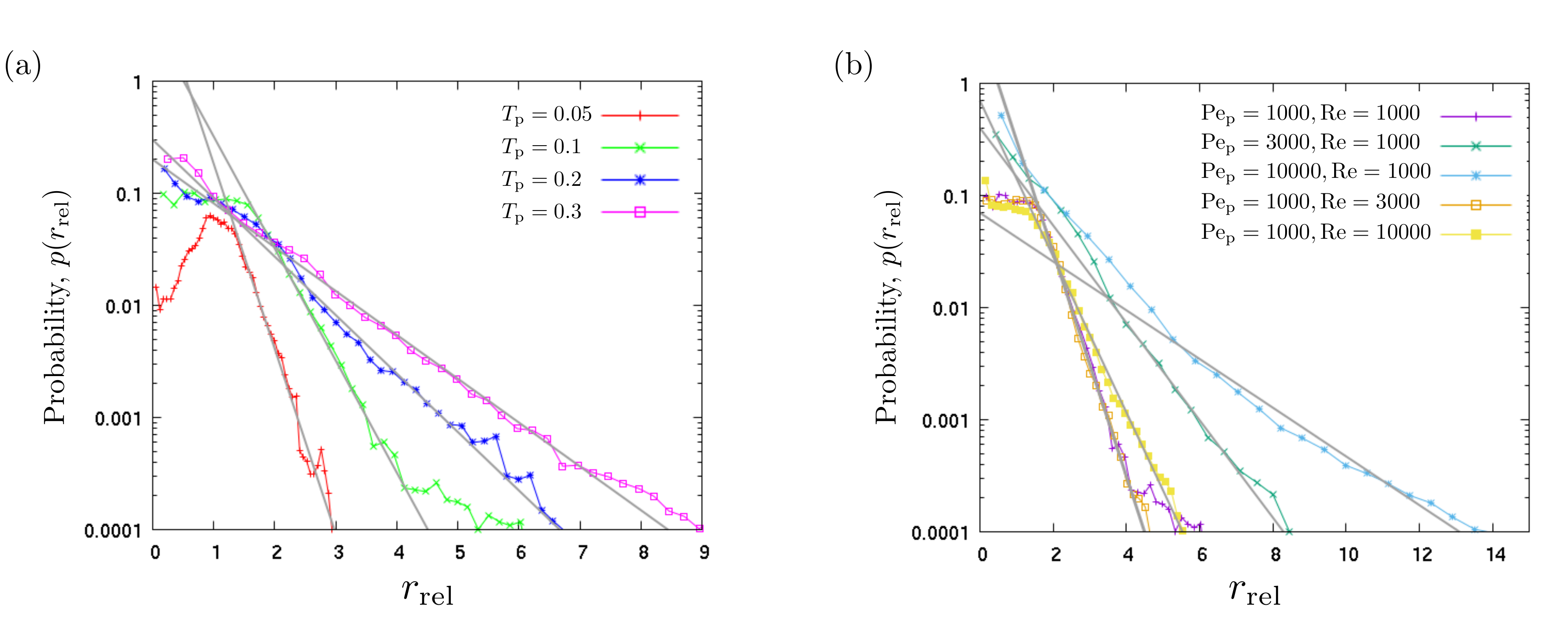}}
\end{figure}

 To explore the properties of this exponential tail, we present PDFs of $r_{\rm{rel}}$ taken during the peak of the mixing event for different simulations at fixed $\Reyn = 1000$ and $\Pep = 1000$ for varying $T_{\rm{p}}$ in Figure \ref{pdf_compare_Pep_Tp}a. We observe that as $T_{\rm{p}}$ increases, the slope of the exponential tail becomes shallower as the maximum value of $r_{\rm{rel}}$ achieved in the simulation increases. In Figure \ref{pdf_compare_Pep_Tp}b, we present PDFs of $r_{\rm{rel}}$ for varying $\Reyn$ and $\Pep$ at fixed $\Tp = 0.1$, taken again at the maximum of the mixing event. We see that the tail widens with increasing $\Pep$ but not with $\Reyn$, which is consistent with our finding that $\Reyn$ does not directly influence the maximum particle concentration achievable (at these parameter values and in this model), but $\Pep$ on the other hand does.

\begin{figure} 
\caption{The slope of the exponential tail $b$ as a function of $\urmstot^2 T_{\rm{p}} {\rm{Pep}}$ for simulations at various $\Tp$, $\Reyn$, and $\Pep$. In all cases, the slope of the PDF is measured during the peak of the mixing event.}
\centering  
\label{exp_dat} 
\vspace{.3cm}
{\includegraphics[width=0.5\pdfpagewidth]{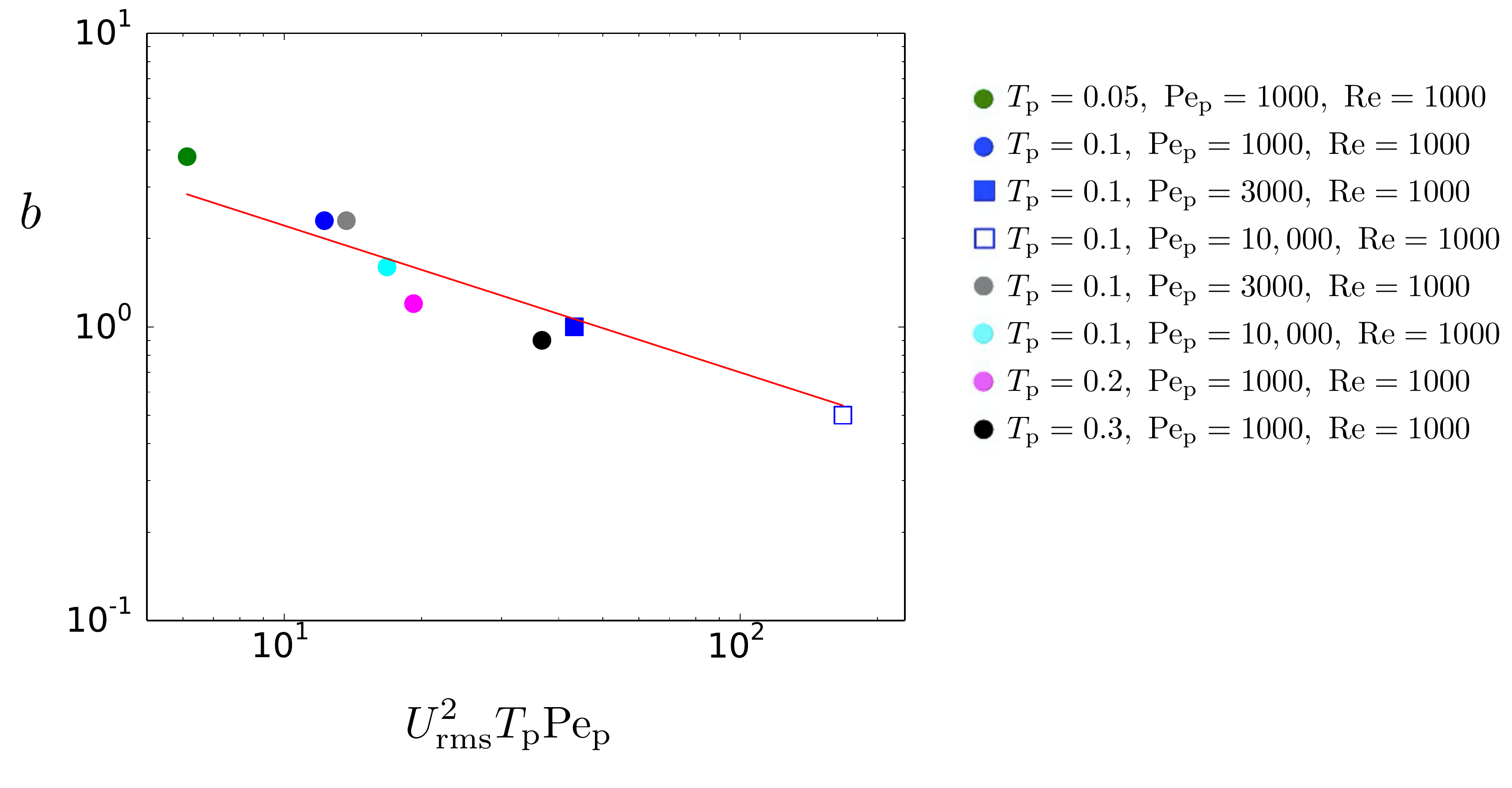}}
\end{figure}

We have fitted an exponential function $f(x) = e^{-bx}$ to the tail of the PDF for each of the cases described above. Figure \ref{exp_dat} shows $b$ as a function of $\urmstot^2 \Tp \Pep$ (where $\urmstot$ and $b$ are computed at the same times). We find that $b \sim [\urmstot^2 \Tp \Pep]^{-1/2}$, which is the same scaling for $(r'/\bar{r})_{\rm{rms}}^{-1}$.  This is perhaps not a coincidence, since the rms of $r_{\rm{rel}} = 1+ r' / \bar{r}$ would be equal to $1/b$ if the distribution was exactly exponential with slope $b$. 

\section{Summary, applications, and discussion}\label{sec:summary} 
\subsection{Summary}\label{subsec:summary}
In this work, we studied preferential concentration in a two-way coupled particle-laden flow subject to the particle-driven convective instability, using DNSs of the two-fluid equations. We constructed an estimate of the typical turbulent eddy velocity in the mixing event as $\urms = \sqrt{r_0 g \sigma}$ (written here dimensionally), where $r_0$ is the ratio of the typical particle mass density excess in the layer to the fluid density, $g$ is gravity, and $\sigma$ is the unstable layer height. Using this, we then constructed an estimate of the particle Stokes number as $\Tp = \tau_p (r_0 g/ \sigma)^{1/2}$, where $\tau_p$ is the dimensional particle stopping time. We found that for $\Tp \le 0.01$, the system properties are indistinguishable from those obtained using the equilibrium Eulerian formalism, while for $\Tp \ge 0.01$, preferential concentration can cause an increase in the particle density in regions of low vorticity or high strain rate, as predicted by \citep{maxey1987gravitational}. The maximum particle concentration enhancement over the local mean,  $\max(\rho_p' / \bar{\rho}_p)$, can be predicted from simple arguments of dominant balance to scale as $\max(\rho_p' / \bar{\rho}_p) \sim \urms^2 \tau_p / \kappa_p$, where $\kappa_p$ is the dimensional particle diffusivity used in the two-fluid model. We verified that this scaling holds for a range of simulations with varying input parameters, as long as $\Tp > 0.01$, and $ \urms^2 \tau_p / \kappa_p > 1$. In this regime, we also found that the probability distribution function of the quantity $\rho'_p / \bar{\rho}_p$ has a root mean square value that scales as $(\urms^2 \tau_p / \kappa_p)^{1/2}$ and an exponential tail whose slope scales as $(\urms^2 \tau_p / \kappa_p)^{-1/2}$.

\subsection{Applications}\label{subsec:applications}
We can use the model proposed in Section \ref{sec:analysis} to predict the maximum particle concentration enhancement over the mean for several applications, where the main source of turbulence is the particle-driven convective instability. We first look at ash created by volcanic eruptions,  droplets in stratus clouds, and sediments suspended in turbidity currents. In all these cases, the particle stopping time is given by 
\begin{equation}\label{eqn:stoppingtime}
	\tau_p = \frac{m_p}{6\pi r_p \rho_f \nu},
\end{equation}
where $r_p$ is the particle radius, and so, the terminal settling velocity is given by
\begin{equation}
	w_s =  \tau_p \bigg(\frac{\rho_s - \rho_f}{\rho_s}\bigg) g = \frac{2}{9} \bigg(\frac{\rho_s - \rho_f}{\rho_f} \bigg) \frac{g}{\nu} r_p^2. \label{eqn:ws}
\end{equation}

Ash particles are generated by volcanic eruptions and have widespread environmental and health implications. Ash particles are transported upwards in the volcanic plume, and eventually spread laterally to form an umbrella cloud in the stratosphere \citep{sparks1976model, woods1995dynamics}. In recent years, there has been renewed interest in predicting the rate of sedimentation of the ash, which is known to depend on preferential concentration \citep{cerminara2016large, carazzo2013particle, webster2012operational}. Suspended ash particles vary widely in radius, especially between the volcanic plume (where $r_p$ ranges from 0.1 mm to 1 mm \citep{harris1983estimating}) and the umbrella cloud (where $r_p$ ranges from 0.1 to 10 $\mu$m, since the larger particles have settled out \citep{carazzo2013particle, webster2012operational}). Similarly, the typical particle concentration $\rho_p$ ranges from $0.1$ $\mu$g/m$^3$ to 1 mg/m$^3$ (see \citep{carazzo2013particle} and references therein) within the umbrella cloud with larger concentration values closer to the eruption site (observed to be $50$ mg/m$^3$ from \citep{harris1983estimating}, for instance). We therefore estimate the Stokes number from \eqref{eqn:stoppingtime} as $\Tp$ given by 
\begin{equation}
	\Tp \approx (2 \times 10^{-7}) \bigg(\frac{\rho_p}{1 \text{ mg/m}^3} \bigg)^{1/2} \bigg( \frac{\sigma}{1 \text{ km}} \bigg) ^{-1/2}  \bigg(\frac{r_p}{10 \text{ }\mu\text{m}} \bigg)^2.
\end{equation}
To arrive at this formula, we have used commonly accepted values for certain parameters, i.e. ($\rho_s-\rho_f)/\rho_f \approx 1000$, $\nu \approx 10^{-5}$ m$^2$/s, and $g \approx 10$ m/s$^2$. We see that for values characteristic of the umbrella cloud, namely $\rho_p$ of order 1 mg/m$^3$, $\sigma$ of order 1 km and $r_p$ of order 10 $\mu$m, $\Tp \sim O(10^{-7}) \ll O(0.1)$. Such a small value of $\Tp$ does not fall in the inertial regime of our model, and thus the effects of preferential concentration due to particle-driven convective instability are negligible. Closer to the volcanic plume, $r_p \sim 0.25$ mm and $\rho_p \sim 50$ mg/m$^3$. Keeping the remaining parameters as before, we find that $\Tp \approx 0.01$, which lies at the boundary of the inertial regime, suggesting that preferential concentration is possible in this case. To determine the maximum particle concentration enhancement, we then use
\begin{equation}\label{eqn:rmax}
	\bigg(\frac{\rho_p'}{\bar{\rho}_p}\bigg)_{\max} =  \frac{1}{4}\frac{| \mathbf{u} | ^2 \tau_p}{\kappa_p} = 25 \bigg(\frac{\rho_p}{1 \text{ mg/m}^3} \bigg) \bigg( \frac{\sigma}{1 \text{ km}} \bigg)  \bigg(\frac{r_p}{10 \text{ }\mu\text{m}} \bigg)^{-1}
\end{equation}
where $|\mathbf{u}|$ is calculated from \eqref{eqn:u} and $\kappa_p\sim w_s r_p$  (see \citep{ham1988hindered, nicolai1995particle, segre2001effective}). 
Thus, from equation \eqref{eqn:rmax} for conditions closer to the volcano with  $\sim$0.25 mm ash particle and $\rho_p \sim 50$ mg/m$^3$, we obtain $(\rho_p' / \bar{\rho_p} )_ {\max} \approx O(100)$, and so, the inertial concentration mechanism may be important in this case.

We also considered other geophysical applications in which particle-driven convection  could be relevant, such as stratus clouds and turbidity currents. Using commonly accepted values for these systems, we found that the estimated Stokes number $\Tp$ is always very small, and therefore does not fall under the inertial regime where preferential concentration takes place (see Appendix \ref{app:geo} for details).

A more interesting application of our model can be found in the astrophysical context of a collapsing protostar,  i.e. a contracting cloud composed of a mixture of gas and dust particles that will eventually lead to the formation of a star. The contraction is usually slow and quasi-hydrostatic, and the gas is generally stably stratified. However, we expect that waves or shocks propagating through it would create inhomogeneities in the dust concentration, that are conceivably gravitationally unstable to particle-driven convective instabilities. With this in mind, we consider typical interstellar dust particles to have a radius of size  $r_p \sim 10$ $\mu$m and solid density $\rho_s \sim O(10^3)$ kg/m$^3$. The gas density within a cloud of radius $R$ astronomical units (AU, where 1 AU = $10^{11}$ m) is typically of order $\rho_f \sim O(10^{-12})$ kg/m$^3$. The dust-to-gas mass ratio in these clouds is of order $r_0 \sim 0.01$, and we anticipate large-scale perturbations above this mean value driven by waves or shocks to be of the same order of magnitude.

Given that the size of the dust particles in this case is much smaller than the mean free path of the gas, the stopping time is now given by 
\begin{equation}\label{eqn:taupAstro}
	\tau_p = \frac{r_p \rho_s}{c \rho_f},
\end{equation}
where $c$ is the sound speed (i.e. $c \approx k_B T/m_H$, where $k_B = 1.38 \times 10^{-23}$ m$^2$ kg s$^{-2}$ K$^{-1}$ is the Boltzmann constant, $m_H \approx 10^{-27}$ kg is the mass of a hydrogen molecule, and $T$ is the local temperature, which is of the order 10 K in clouds \citep{tobin20120}). Using $g = GM_\star / R^2$ in \eqref{eqn:taupAstro}, where $G = 6.7 \times 10^{-11}$ m$^3$ kg$^{-1}$ s$^{-2}$ is the gravitational constant and $M_\star$ is the mass of the core of the protostar, we then find that the non-dimensional stopping time is given by 
\begin{align*}
	T_p = (10^{-1}) \bigg(\frac{r_p}{10\text{ }\mu\text{m}}\bigg) &\bigg(\frac{\rho_s}{10^3 \text{ kg/m}^3}\bigg) \bigg(\frac{\rho_f}{10^{-12}\text{ kg/m}^3}\bigg)^{-1}\\ &\bigg(\frac{T}{10 \text{ K}}\bigg)^{-1/2} \bigg(\frac{r_0}{0.01}\bigg)^{1/2} \bigg( \frac{M_\star}{M_\odot} \bigg)^{1/2} \bigg(\frac{R}{100 \text{ AU}}\bigg)^{-1}\bigg( \frac{\sigma}{0.01\text{ AU}}\bigg)^{-1/2},
\end{align*}
where $M_\odot = 2 \times 10^{30}$ kg is the mass of the Sun. Here, we see that by using typical values for a protostar and assuming that the particle density inhomogeneities are initially of size $0.01$ AU, then $\Tp$ lies within the inertial regime. The relative maximum particle concentration can be then written as
\begin{align*}
	\bigg(\frac{r'}{\bar{r}}\bigg)_{\max} = (10^{11})\bigg(\frac{r_p}{10\text{ }\mu\text{m}}\bigg)^3 \bigg(\frac{\rho_s}{10^{3} \text{ kg/m}^3}\bigg)\bigg(\frac{r_0}{0.01}\bigg) &\bigg(\frac{T}{10 \text{ K}}\bigg)^{-1}\\ &\bigg(\frac{M_*}{M_\odot}\bigg) \bigg(\frac{R}{100 \text{ AU}}\bigg)^{-2} \bigg(\frac{\sigma}{0.01 \text{ AU}}\bigg).
\end{align*}
While this relative enhancement is huge, it is not sufficient to bring particles in contact with one another. Indeed, the associated volume fraction of particles would be $\Phi' = r_0(\rho_f/\rho_s)(r'/\bar{r})_{\max} \approx 10^{-6}$. Nevertheless, this does imply that the particle collision rate within these enhanced regions would dramatically increase, suggesting that preferential concentration due to particle-driven convective instabilities could play a role in star and planet formation. 

\subsection{Discussion}
Assuming that the model described in Section \ref{sec:analysis} and summarized in \ref{subsec:summary} is generally valid in particle-laden turbulent flows, it provides a very simple way of estimating the expected enhancement in the local particle density due to preferential concentration, which could be very useful for predicting its impact on other processes, such as particle growth or enhanced settling, as demonstrated in \ref{subsec:applications}. However, several caveats of the model need to be kept in mind before doing so. First and foremost is the fact that the maximum particle concentration enhancement over the local mean depends explicitly on the particle diffusivity $\kappa_p$, which is derived from a simplistic model of the interaction between the particles and the fluid, as well as among the particles themselves. In the limit where Brownian motion is the dominant contribution to the particle diffusivity, then the model is likely to be valid. This is the case for instance in astrophysical applications. However, when the interaction of the particle with its own wake or with the wakes of other particles dominates, then the simple diffusion model $\kappa_p \nabla^2 \rho_p$ presumably fails to capture some of their more subtle consequences and should only be used with considerable caution. Comparisons of the model with particle-resolving simulations will help elucidate whether any of our results still holds for more realistic situations.  

Another caveat of the model is the fact that it has only been validated so far in moderately turbulent flows, for which the inertial range is fairly limited. In more turbulent systems, where the inertial range spans many orders of magnitude in scales, the Stokes number at the injection scale could be quite different from the Stokes number at the Kolmogorov scale. Assuming a Kolmogorov power spectrum for the kinetic energy, for instance, it is easy to show that the Stokes number increases weakly with wavenumber, and can be substantially larger at the Kolmogorov scale than at the injection scale when the Reynolds number is very large. This raises the question of whether the model remains applicable when this is the case.  Finally, we note that the model has so far only been tested in the context of particle-driven convection, where the two-way coupling between the particles and the fluid likely influence the turbulent cascade. It remains to be determined whether the same scalings are found in flows where the source of the turbulence is independent of the particles (such as mechanically driven turbulence, or thermal convection, for instance). If this is the case, our findings may have further implications for engineering or geophysical flows. Both of these questions will be the subject of future work.

There are also several other questions that remain to be answered. The simulations presented in Section \ref{subsec:PepRey}, for instance, clearly show that the particle P\'eclet number influences the typical width and separation of the regions of high particle density, but this effect remains to be explained and modeled. This will require a better understanding of the influence of the two-way coupling between the particles and the fluid on the turbulent energy cascade from the injection scale to the dissipation scale. In particular, it is clear from a cursory inspection of the kinetic energy spectrum (see Figure \ref{spectra_varyPepRe}) that the extent of the inertial range depends equally on the Reynolds number and on the P\'eclet number, suggesting that this two-way coupling dominates the flow dynamics at small scales. Although this is perhaps not surprising, it deserves to be investigated further. Moreover, it would be interesting to see whether the same effect occurs in a system in which the turbulence is not driven by the particles themselves. 

S.N. acknowledges funding by NSF AST-1517927 grant. S.N. was also partially supported by the NSF-MSGI summer program at Lawrence Berkeley Laboratory under the supervision of A. Myers and A. Almgren. Simulations were run on a modified version of the PADDI code, originally written by S. Stellmach, on  the UCSC Hyades cluster and the NERSC Cori supercomputer. The authors thank Eckart Meiburg and Doug Lin for helpful discussions. 

\appendix
\section{Properties of PADDI}\label{app:num_sims}
The governing equations are solved in spectral space using a third-order semi-implicit Adams-Bashforth backward-differencing scheme. Diffusive terms are treated implicitly. Nonlinear terms and drag terms are first computed in real space, then transformed into spectral space using FFTW libraries, and advanced explicitly.  Drag terms are tracked and computed in a way that ensures the total momentum is conserved (other than the dissipation terms) throughout the simulations. 

We encountered various numerical obstacles during the implementation of the two-fluid equations in PADDI-2F that are worth mentioning here. Due to the fact that particle inertia tends to increase particle concentration in certain regions for large enough $\Tp$, one must use a very high spatial resolution to avoid numerical instability. Even when the resolution is large enough to ensure numerical stability, a slight under-resolution can result in the particle concentration being slightly over- or underestimated, resulting in the total mass not being exactly conserved. Indeed, in a spectral code, low resolution can induce the Gibbs phenomenon which can create regions of unphysical negative particle density near the edges of a particle front. In the code, we zero out the negative particle density regions and rescale the particle concentration $r$ at every point in space to ensure that the total particle mass is equal to its initial value at each time step. Note that this ``fix'' is generally not necessary as long as the simulations are well-resolved, but is introduced to reduce errors in the rare occasions where the system does become slightly under-resolved.

\section{Other geophysical applications}\label{app:geo}
We looked at the applicability of our model for the preferential concentration of water droplets found in stratus clouds. These clouds are a more relevant application of our model than convective clouds (i.e. cumulus and cumulonimbus) in which turbulence is primarily driven by thermal convection rather than particle-driven convection. We estimate $r_0$ and $\Tp$ by 
\begin{align}
    &r_0 = (2.5 \times 10^{-4}) \bigg(\frac{\rho_p}{0.25 \text{ g/m}^3} \bigg), \label{eqn2:r0}  \\
	&\Tp \approx (3.5 \times 10^{-6}) \bigg(\frac{\rho_p}{0.25 \text{ g/m}^3} \bigg)^{1/2}  \bigg( \frac{1 \text{ km}}{\sigma} \bigg) ^{1/2}  \bigg(\frac{r_p}{10 \text{ }\mu\text{m}} \bigg)^2,\label{eqn2:Tp}
\end{align}
where $\rho_p$ here is otherwise known as the liquid water content which is typically of the order of 0.25 g/m$^3$ for stratus clouds \citep{frisch2000comparison}. We have also applied commonly accepted values for certain parameters for these formulas (i.e. $\rho_s/\rho_f \approx 1000$, $\nu \approx 10^{-5}$ m$^2$/s, $g \approx 10$ m/s$^2$). According to \eqref{eqn2:Tp}, we see that for any reasonable droplet size, $\Tp$ is in the regime where preferential concentration would not occur due to the particle-driven convective instability.

We now look at particle concentration in the context of turbidity currents which play a vital role in the global sediment cycle. We consider sediments consisting of clay, silt, or sand that vary in radius from $O(10^{-4}) - O(10^{-1})$ cm (where clay is found at the lower end of this range, while sand particles are found at the larger end) with solid density typically around $\rho_s \approx 2000$ kg/m$^{3}$. For a particle volume fraction $\Phi$ in the dilute regime, $\Phi \lesssim 0.01$ and so, $r_0 \lessapprox 0.02$, and
\begin{equation}
	\Tp \approx (2 \times 10^{-4}) \bigg( \frac{10 \text{ m}}{\sigma} \bigg) ^{1/2}  \bigg(\frac{r_p}{0.1 \text{ mm}} \bigg)^2 \bigg( \frac{\Phi}{0.01} \bigg) ^{1/2},
\end{equation}
in which we have assumed that $(\rho_s - \rho_f)/\rho_f \sim O(1)$. We therefore see that even for the largest particle size and for the maximum volume fraction allowable, for any reasonable value of $\sigma$, $\Tp \ll 0.1$ so preferential concentration due to particle-driven convective instabilities is again negligible. 

\begin{table}
\caption{Terms defined from text.} \label{table:defns} % title of Table
\centering % used for centering table
\begin{tabular}{l l} % centered columns (4 columns)
\hline\hline %inserts double horizontal lines
Definition & Description  \\ [0.5ex] % inserts table
%heading
  \hline  % inserts single horizontal line
  & \\
					$\bar{r}(z.t) = \overline{r(x,z,t)} = \frac{1}{L_x} \int r(x,z,t)$				&	 \multicolumn{1}{m{7cm}}{Horizontal average of the particle concentration at a given height at time $t$.}			\\
					$\rmax (z,t) = \max_x r(x,z,t)$				&   \multicolumn{1}{m{7cm}}{Maximum value of the particle concentration at a given height at time $t$.}      		\\
					$\rrms (z,t) = \bigg[ \overline{[r(x,z,t) - \bar{r}(z,t)]^2} \bigg]^{1/2}	$		&   \multicolumn{1}{m{7cm}}{Typical enhancement over $\bar{r}$ at a given height at time $t$.}         \\
					$\urms (z,t) = [\overline{u(x,z,t)^2}]^{1/2} $ &  \multicolumn{1}{m{7cm}}{Root mean square of the $x-$component of the fluid velocity at a given height at time $t$.}    	 \\ 
					$r_{\rm{rel}}(x, z, t) = { \frac{r(x,z,t)}{\bar{r}(z,t)}  }$ &  \multicolumn{1}{m{7cm}}{Relative particle concentration at time $t$.}    	 \\ 
					& \\
				    \hline 
				    & \\
				   Extracted in the bulk of the particle layer: & \\ 
					$z_{\max}(t)$			   &  \multicolumn{1}{m{7cm}}{Height corresponding to the maximum value of $\bar{r}$ at time $t$.}   \\
					$\bar{r}^*(t) = \bar{r} (z_{\max}, t)$				& \multicolumn{1}{m{7cm}}{Maximum value of $\bar{r}$ at time $t$.}		         	 	\\
					$\urms^*(t) = \urms(z_{\max}, t) $				&  \multicolumn{1}{m{7cm}}{Value of $\urms$ measured at $z_{\max}$ at time $t$.}       	   \\
					& \\
				    \hline 
				    & \\
					$\rsup(t) = \max_{x,z}  {r(x,z,t)} $					& \multicolumn{1}{m{7cm}}{Maximum particle concentration in the domain at time $t$.}   				 		\\
					$\usup(t) = \max_{x,z} u(x,z,t) $					&		\multicolumn{1}{m{7cm}}{Maximum value of the horizontal velocity of the fluid at time $t$.}  					\\ 
					& \\
\hline %inserts single line
\end{tabular}
\label{table:nonlin} % is used to refer this table in the text
\end{table}
%
%
%% Create the reference section using BibTeX:
%%\bibliographystyle{jfm}
%%% Note the spaces between the initials
%%\bibliography{jfm-instructions}
%
%
%% Create the reference section using bbl (not BibTex):

\end{document}